\documentclass[12pt]{article}
\pdfoutput=1
\usepackage[centertags]{amsmath}
\allowdisplaybreaks[1]
\usepackage{array,multirow}
\numberwithin{equation}{section}
\usepackage{amssymb}
\usepackage{float}
\usepackage{graphicx}
\usepackage{color}
\usepackage{relsize}
\usepackage{verbatim}
\usepackage{mathtools}

\usepackage{epsfig}

\def\Or[#1]{{\text{O}}\left({#1}\right)}
\def\dotl[#1,#2]{\left\langle #1, #2 \right\rangle}
\def\dotlb[#1,#2]{[ #1, #2 ]}
\def\dotp[#1,#2]{(#1) \cdot (#2)}
\def\aff[#1,#2]{\hat{#1}(#2)}
\def\n4sym{{\cal N}=4 SYM}
\def\>{\rangle}
\def\<{\langle}
\def\weight[#1,#2,#3]{\{(#1),#2,#3\}}
\def\ads[#1]{$\text{AdS}_{#1}$}

\newcommand{\ba}{\begin{eqnarray}}
\newcommand{\ea}{\end{eqnarray}}
\newcommand{\be}{\begin{eqnarray}}
\newcommand{\ee}{\end{eqnarray}}
\newcommand{\bq}{\begin{equation}}
\newcommand{\eq}{\end{equation}}
\newcommand{\benn}{\begin{equation*}}
\newcommand{\eenn}{\end{equation*}}
\newcommand{\bi}{\begin{itemize}}  
\newcommand{\ei}{\end{itemize}}

\newcommand{\CD}{{\cal D}}

\newcommand{\CL}{{\cal L}}

\newcommand{\CO}{{\cal O}}

\newcommand{\nn}{\nonumber}

\newcommand\oo\infty
\newcommand\s\sigma
\newcommand\de\delta
\newcommand\De\Delta

\newcommand\f\phi
\newcommand\g\gamma
\newcommand\x\times

\usepackage{jheppub}

\makeatletter
\def\@fpheader{\vspace{-.1cm}}
\makeatother

\title{An Exact Operator That Knows Its Location}

\author[a]{N. Anand,}
\author[a]{Hongbin Chen,}
\author[b]{A. Liam Fitzpatrick,} 
\author[a,c,d]{Jared Kaplan,}
\author[a]{Daliang Li}
\affiliation[a]{Department of Physics and Astronomy,  Johns Hopkins University, \\
Charles Street, Baltimore, MD 21218, USA} 
\affiliation[b]{Department of Physics, Boston University, \\
Commonwealth Avenue, Boston, MA 02215, USA}
\affiliation[c]{Center for Quantum Mathematics and Physics (QMAP) \\
University of California, Davis, California 95616, USA}
\affiliation[d]{Stanford Institute for Theoretical Physics \\
Stanford University, Palo Alto, CA 94305, USA
}

\abstract{ 
We use conformal symmetry to define an AdS$_3$ proto-field $\phi$ as an exact linear combination of Virasoro descendants of a CFT$_2$ primary operator $\CO$.  We find that both symmetry considerations and a gravitational Wilson line formalism lead to the same results.  The operator $\phi$ has many desirable properties; in particular it has correlators that agree with gravitational perturbation theory when expanded at large $c$, and that automatically take the correct form in all vacuum AdS$_3$ geometries, including BTZ black hole backgrounds.  In the future it should be possible to use $\phi$ to probe bulk locality and black hole horizons at a non-perturbative level.
}   

\arxivnumber{}

\begin{document}
   
\maketitle
\flushbottom
 
\section{Introduction}

To resolve the black hole information paradox in AdS/CFT, we must understand how to describe local AdS dynamics in terms of CFT data and observables.  Unfortunately, bulk gauge redundancies could render AdS reconstruction ambiguous, and the existence of black holes at high-energies suggests that local physics may not be well-defined.  We will argue that the Virasoro symmetry of CFT$_2$ provides a sort of beachhead into AdS$_3$, making it possible to exactly define a bulk `proto-field' $\phi$ as a specific linear combination of Virasoro descendants of a given local primary operator $\CO$.

The simplest AdS/CFT  observable is the vacuum bulk-boundary correlator
\be
\< \phi(X) \CO(P) \> = \frac{1}{(P \cdot X)^{\Delta}},
\ee
which is determined by conformal symmetry up to an overall constant.  From this correlator alone one can derive a formula for a proto-field $\phi(X)$ as a linear combination of global conformal descendants of the primary operator $\CO$ \cite{Banks:1998dd, Hamilton:2005ju, Miyaji:2015fia,Nakayama:2015mva}.   At this level, bulk reconstruction is purely kinematical, following entirely from the assumption that conformal transformations act on $\phi$ as AdS isometries.  

In the case of AdS$_3$/CFT$_2$, Virasoro conformal transformations act as asymptotic symmetries.  So it is natural to expect that the bulk-boundary correlator should be uniquely determined in any geometry that can be related to the vacuum by a Virasoro symmetry.  In rather different words, we expect that all correlators of the form
\be
\label{eq:phiOTTb}
\< \phi(X) \CO(z,\bar{z}) T(z_1) \cdots T(z_n) \bar T(\bar w_1) \cdots \bar T(\bar w_m) \> 
\ee
can be determined by symmetry once we fix a gauge for the bulk gravitational field.  This leads to a unique expression for a Virasoro proto-field operator $\phi(X)$ as a linear combination of Virasoro descendants of the CFT$_2$ primary $\CO$.  These proto-field operators will automatically `know' about the bulk geometry associated with heavy distant sources, meaning that they perform bulk reconstruction at an operator level.  In this paper we will explain how to identify and explicitly compute $\phi(X)$ as a CFT$_2$ operator.  
We will be led to the potentially surprising conclusion that an exact (non-perturbative in $c$) condition uniquely determines $\phi$ in our Fefferman-Graham type gauge.

We will determine $\phi(X)$ in two distinct but ultimately equivalent ways.  The first is based on an extension of gravitational Wilson lines \cite{Verlinde:1989ua, KrausBlocks, deBoer:2013vca, Ammon:2013hba, Besken:2016ooo, Fitzpatrick:2016mtp, Besken:2017fsj} as OPE blocks \cite{Czech:2016xec}.  We will introduce a `bulk-boundary OPE block' that encapsulates the projection of the (non-local) operator $\phi(X) \CO(x)$ onto the vacuum sector.  This provides an explicit prescription for computing all correlators of the form of equation (\ref{eq:phiOTTb}).  Our second method is based purely on imposing Virasoro symmetry, resulting in a very simple, non-perturbative definition for $\phi(X)$.  This also makes it possible to determine the correlators of equation (\ref{eq:phiOTTb}) via a simple recursion relation.  The  proto-field operator that we will obtain has a number of desirable properties:
\begin{itemize}
\item Virasoro transformations act on the scalar field $\phi(X)$ as infinitesimal bulk diffeomorphisms preserving the gauge.  At the semiclassical level, $\phi(X)$ obeys the Klein-Gordon equation in any  vacuum geometry.
\item Correlators of $\phi$ with stress tensors are causal and have only those singularities dictated by the gravitational constraints \cite{Kabat:2011rz, Kabat:2012av, Kabat:2013wga}, matching bulk perturbation theory.  Correlators of $\phi(X)$ reduce to those of $\CO(x)$ when we extrapolate $\phi(X)$ to the boundary. Equation (\ref{eq:phiOTTb}) reduces to $\< \CO \CO T \cdots \bar T \cdots \>$; in fact there is a simple recursion relation that computes vacuum correlators, generalizing well-known relations \cite{Ginsparg} for correlators of CFT$_2$ primaries with stress tensors.  
\end{itemize}
With our exact definition for $\phi(X)$, it is possible to study the impact of non-perturbative gravitational effects on bulk observables.  This means that one could study $\phi(X) \phi(Y)$ at short distances, and directly probe near black hole horizons without relying on bulk perturbation theory.   

There is a large literature on bulk reconstruction in AdS/CFT employing a variety of philosophies and methods, 
for example \cite{Banks:1998dd, Bena:1999jv, Hamilton:2005ju, Hamilton:2006az, Kabat:2011rz, Kabat:2012av, Kabat:2013wga, Heemskerk:2012mn, Papadodimas:2012aq, Kabat:2015swa, Nakayama:2015mva, Guica:2015zpf, Kabat:2016zzr, Czech:2016xec, Nakayama:2016xvw, Faulkner:2017vdd, Almheiri:2017fbd,Verlinde:2015qfa,Lewkowycz:2016ukf}.\footnote{We believe the proposal in \cite{Verlinde:2015qfa,Lewkowycz:2016ukf} is different from ours.}  The most common approach expresses bulk fields in terms of local CFT operators integrated against a kernel \cite{Banks:1998dd, Hamilton:2005ju, Hamilton:2006az}.  We will take a somewhat different approach  \cite{Paulos:2016fap, daCunha:2016crm, Guica:2016pid}; our  scalar operator $\phi(y, 0,0)$ will be expressed in a boundary operator expansion\footnote{The idea of performing bulk reconstruction using a boundary operator expansion was briefly discussed in \cite{Paulos:2016fap}.  The global  AdS results have been worked out by M. Paulos in unpublished work.  Note that the boundary operator expansion appears local on the boundary, but due to the infinite sum it should really be viewed as a non-local CFT operator, for the same reason that $e^{x \partial} \chi(0) = \chi(x)$ should not be viewed as a local operator at the origin.}  (BOE) \cite{Cardy:1991tv}
\be
\label{eq:BOEforphi}
\phi(y,0,0) = \sum_{N=0}^\infty \lambda_N  y^{2h+2N} \CL_{-N} \bar \CL_{-N} \CO(0)
\ee
where $\CL_{-N}$ and $\bar \CL_{-N}$ are linear combinations of products of Virasoro generators at level $N$, and $\lambda_N = \frac{(-1)^N}{N! (2h)_N}$. In the global limit ($c\rightarrow\infty$), we have $\lim_{c\rightarrow\infty}\CL_{-N} = L_{-1}^N$. At finite $c$, we will show that $\mathcal{L}_{-N}\mathcal{O}$ satisfies the bulk primary conditions
 \be
 L_{m} \mathcal{L}_{-N}\mathcal{O}=0, \qquad \text{for } m\ge2.
\ee
and similarly for $\bar \CL_{-N} \CO$.  Roughly speaking, these conditions say that $\phi$ is as primary as it can be and still move around under AdS bulk isometries.
 In the smearing function language, we are computing $\phi$ as an infinite sum of operators\footnote{As was shown by Kabat and Lifschytz \cite{Kabat:2012av, Kabat:2013wga}, because of the gravitational gauge constraints $\phi$ must include contributions from the scalar descendants of quasi-primaries with non-zero spin, such as $\partial^\mu \partial^\nu [T_{\mu \nu}\CO]$, even though $\phi$ itself is a bulk scalar field.  Thus it's not entirely clear how smearing functions can be used to describe our results. } of the schematic form $\CO,  [T \bar \partial^2 \CO], \cdots , [T \partial^2T \bar T \bar \partial^4 \CO]$, $\cdots$, though we will not express our results in this way.

The outline of this paper is as follows.  In section \ref{sec:GravitationalWilsonLines} we explain the bulk-boundary OPE block idea, and then show how the vacuum $\phi(X) \CO(z)$ OPE block can be derived using gravitational or Chern-Simons Wilson lines.  We begin   section \ref{sec:SymmetryDefinitionPhi} by providing an exact algebraic definition for $\phi$ compatible with the results of section \ref{sec:GravitationalWilsonLines}.  Then we show that this simple definition follows from considerations of symmetry.  We solve for $\phi$ explicitly in various cases, and then show how our definition leads to new recursion relations for correlators of $\phi$ with boundary stress tensors.  We collect various technical results and background material in the appendices.  Appendix \ref{app:GlobalBOEBasics} may be  useful for readers who are most familiar with the HKLL \cite{Hamilton:2005ju} smearing procedure, and want to understand how our approach, in the simple global conformal case, can be reduced to theirs.  All formulas in this paper are written in Euclidean signature.

\section{Bulk Reconstruction from Gravitational Wilson Lines}
\label{sec:GravitationalWilsonLines}

The operator product expansion (OPE) expresses a product of separated local operators $\CO_1(x_1) \CO_2(x_2)$ as an infinite sum of local operators at a single point.  It is very natural to gather the contributions to the OPE that come from a single conformal primary and its descendants.  This has been dubbed \cite{Czech:2016xec} an `OPE block'.   In the case of CFT$_2$, the Virasoro OPE blocks can be computed using Chern-Simons Wilson lines \cite{Fitzpatrick:2016mtp}.

\begin{figure}
\centering
\includegraphics[width=0.45\textwidth]{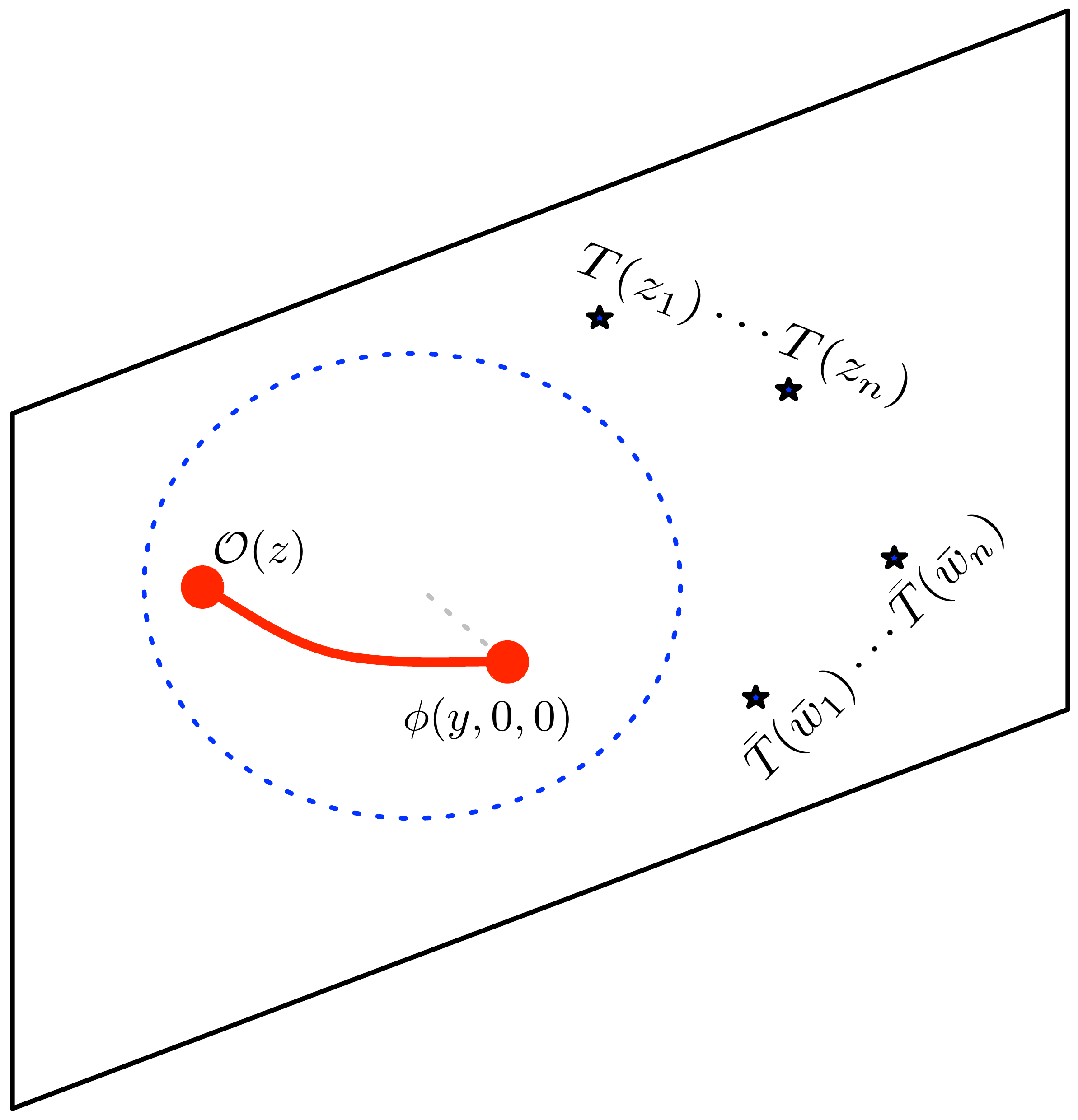}	
\caption{This figure portrays  a bulk-boundary OPE block used to compute the correlator (\ref{eq:phiOTTb}).  The red line denotes the gravitational or Chern-Simons Wilson line, while the blue circle suggests radial quantization around the block, so that it creates a definite linear combination of Virasoro descendants of the identity.  The explicit computation involves line integrals over stress tensor correlators.}
\label{fig:BulkBoundaryOPEBlock}
\end{figure}

In this work we will be studying an AdS$_3$ proto-field $\phi(X)$ as a CFT$_2$ operator, and we focus on Euclidean signature.  Although $\phi(X)$ may be somewhat non-local, on the border of a sufficiently large region in the CFT containing $\phi(X)$, we expect that it should still be possible to perform a radial quantization, as shown in figure \ref{fig:BulkBoundaryOPEBlock}.  This suggests that we can study OPE blocks involving $\phi(X)$ and other operators.  We will be focusing on the simplest such object, the scalar Virasoro vacuum OPE block 
\be
\label{eq:BulkBoundaryOPEBlock}
\phi(y, 0, 0) \CO(z,\bar z) = \frac{y^{2h}}{(y^2 + z \bar z)^{2h}} + \cdots
\ee
where the ellipsis denotes non-identity Virasoro generators (e.g. $L_{-6} \bar L_{-4}^2$) with coordinate-dependent coefficients, and we have labeled $\phi$ using the coordinates of the AdS$_3$ vacuum metric
\be
ds^2 = \frac{dy^2 + dz d \bar z}{y^2}
\ee
Note that we have already identified the contribution of the identity operator in equation (\ref{eq:BulkBoundaryOPEBlock}) as the vacuum correlator $\< \phi \CO \>$, which is fixed by conformal symmetry.  All of the remaining terms in equation (\ref{eq:BulkBoundaryOPEBlock}) would be fixed if we knew all correlators of the form (\ref{eq:phiOTTb}), because the Virasoro generators are just the modes in an expansion of the stress tensors $T(z)$ and $\bar T(\bar z)$.

Building on prior work \cite{Fitzpatrick:2016mtp}, we will make the following proposal for the $\phi \CO$ OPE block.  The general asymptotically AdS$_3$ vacuum metric can be written as \cite{Banados:1998gg, Roberts:2012aq}
\be
\label{eq:MetricwithT}
ds^2 = \frac{dy^2 + dz d \bar z}{y^2} - \frac{6 T(z)}{c} dz^2   - \frac{6 \bar T(\bar z)}{c} d \bar z^2 + y^2 \frac{36 T(z) \bar T( \bar z)}{c^2} dz \bar dz
\ee
This amounts to a choice of gauge for the bulk gravitational field.  Normally the objects $T(z)$ and $\bar T( \bar z)$ appearing in this equation are viewed as classical functions, but let us instead view them as CFT$_2$ stress tensor operators.  We define the bulk-boundary OPE block as the operator defined by the propagation of a (quantum, first-quantized) particle from the location of $\CO$ on the boundary to that of $\phi$ in the bulk.  Formally, this means that the bulk-boundary OPE block can be thought of  as a world-line path integral
\be
\label{eq:OPEBlockAsPathIntegral}
\phi(X) \CO(0)|_{\text{vac}} = \int \CD Y(\tau) \, e^{- m \int_0^X d\tau\sqrt{g_{\mu \nu} \dot Y^\mu \dot Y^\nu}} ,
\ee
where on the LHS we are restricting to the vacuum sector of the operator product. On the RHS we interpret $g_{\mu \nu}$ as a quantum operator dependent on $T, \bar T$ as defined in equation (\ref{eq:MetricwithT}), and  $Y^\mu(\tau)$ is world-line connecting $\phi$ and $\CO$.  The mass $m$ of the particle will be related to the dimension of $\CO$ by $m^2 = 2h (2h - 2)$.  Equation (\ref{eq:OPEBlockAsPathIntegral}) defines the OPE block as an infinite sum of products of line integrals of the CFT$_2$ stress tensor.  We have sketched the OPE block in figure \ref{fig:BulkBoundaryOPEBlock}.

In a certain sense, we will use equation (\ref{eq:OPEBlockAsPathIntegral}) more for conceptual purposes than for computation ones.  To use (\ref{eq:OPEBlockAsPathIntegral}) directly would require defining the path integral measure precisely; in practice, we will circumvent this kind of issue by starting with the exact CFT result for (\ref{eq:OPEBlockAsPathIntegral}) in the AdS vacuum and uplifting to nonzero $T,\bar{T}$ by performing diffeomorphisms, as we describe below.  Nevertheless, it is useful to bear equation (\ref{eq:OPEBlockAsPathIntegral}) in mind as it intuitively captures what we are trying to achieve in defining $\phi \CO$, and furthermore it should agree with our practical definition in a semiclassical limit where ambiguities in the path integral measure do not arise.  So when we compute the bulk-boundary OPE block in the presence of operators with dimensions $h_H \propto c$ at large $c$, then we can approximate $\phi \CO$ by including only the semiclassical expectation value $\< T(z) \> \propto \frac{h_H}{c}$.  This immediately leads to the correct $\phi \CO$ correlators in a semiclassical background, such as that of a BTZ black hole.   Relatedly, our  prescription will also lead to a $\phi(X)$ that satisfies the Klein-Gordon equation in the semiclassical metric of equation (\ref{eq:MetricwithT}).  We review this elementary fact in appendix \ref{app:KleinGordon}. We also provide a more detailed discussion of (\ref{eq:OPEBlockAsPathIntegral}) and its regulation in appendix \ref{subsec:FromClassicalBackgroundtoMultiTVacuumCorrelators}.   

In the remainder of this section we will use equation (\ref{eq:OPEBlockAsPathIntegral}) to explicitly compute various correlation functions, and demonstrate that  the results  reduce to those of \cite{Fitzpatrick:2016mtp} when we take $\phi$ to the boundary.  In fact we will find that we can  reformulate equation (\ref{eq:OPEBlockAsPathIntegral}) in terms of sl$(2)$ Wilson lines as 
\be
 \phi(y,z_2, \bar{z}_2) \CO(z_1, \bar{z}_1)|_{vac} = \left. P \left\{ e^{ \int_{z_1}^{z_2} dz A_z+ \int_{\bar{z}_1}^{\bar{z}_2} d\bar{z} \bar{A}_{\bar{z}}} \right\} \left( \frac{y}{y^2 + x \bar{x} }\right)^\Delta \right|_{x=\bar{x}=0}.
 \label{eq:BulkWL}
\ee
As we will explain  in section \ref{sec:ConnectionCSWilson} (where we also define the notation), this is the most natural generalization of prior Chern-Simons Wilson line results \cite{Fitzpatrick:2016mtp} to the case of the bulk-boundary OPE block.  It also makes manifest the fact that as we take $\phi$ to the boundary, we recover the structure of the more conventional $\CO(z_2) \CO(z_1)$ Virasoro OPE block.

\subsection{Computing $\phi(X) \CO(0)$ from a Diffeomorphism}
\label{sec:GravityComputationPhiO}

We will use two facts to formulate an operational definition of equation (\ref{eq:OPEBlockAsPathIntegral}) that can be used for practical computations.  The first is that in pure AdS$_3$, the first-quantized path integral reduces to $e^{-2 h \sigma}$ where $\sigma$ is the (renormalized) length of a geodesic connecting $\CO$ and $\phi$.  The second fact is an explicit diffeomorphism \cite{Roberts:2012aq} that relates metrics of the form (\ref{eq:MetricwithT}) to the pure AdS$_3$ metric.  We will elevate this diffeomorphism to an operator equation, defining a change of coordinates parameterized by a function $f_T(z)$  that maps the  pure AdS$_3$ metric to the operator-valued vacuum metric of equation (\ref{eq:MetricwithT}).  Then we can use the first fact to evaluate the bulk-boundary OPE block as a functional of $f_T(z)$, which itself depends on the operator $T(z)$.  These ideas were inspired by very similar methods that have been used to evaluate Chern-Simons Wilson lines \cite{Fitzpatrick:2016mtp} in order to compute Virasoro OPE blocks; we will see in section \ref{sec:ConnectionCSWilson} that this is not an accident.

The first fact means that in a vacuum metric
\be
ds^2 = \frac{du^2 + d w d \bar w}{u^2},
\ee
we can write  the bulk-boundary correlator as
\be
\label{eq:PureAdSLogOfBulkBoundary}
\phi(u,0,0) \CO(w, \bar w)  = \left( \frac{u}{u^2 + w \bar w} \right)^{2h} .
\ee
In the CFT vacuum, this is an exact CFT result, just the standard scalar bulk-to-boundary propagator that can be derived purely from symmetries of the CFT.  But now we will  generalize it by viewing the coordinates $(u,w,\bar w)$ as the result of an operator valued diffeomorphism from a general vacuum metric of the form of equation (\ref{eq:MetricwithT}).  The diffeomorphism takes the form \cite{Roberts:2012aq}
\be
\label{eq:GeneralVacDiffeomorphism}
w &\to & f(z) - \frac{2 y^2 (f'(z))^2 \bar f''(\bar z)}{4 f'(z) \bar f'(\bar z) + y^2 f''(z) \bar f''(\bar z)}
\nn \\
\bar w &\to & \bar f(\bar z) - \frac{2 y^2 \bar (f'(\bar z))^2  f''( z)}{4 f'(z) \bar f'(\bar z) + y^2 f''(z) \bar f''(\bar z)}
\nn \\
u & \to & y \frac{4 (f'(z) \bar f'(\bar z))^{\frac{3}{2}} }{ 4 f'(z) \bar f'(\bar z) + y^2 f''(z) \bar f''(\bar z) }
\ee
and is parameterized by the independent holomorphic and anti-holomorphic functions $f(z)$ and $\bar f(\bar z)$.  This diffeomorphism has the property that the transformed metric is precisely
\be
ds^2 = \frac{d y^2 + dz d \bar z}{y^2} - \frac{1}{2} S(f,z) dz^2 - \frac{1}{2} S(\bar f, \bar z) d \bar z^2  + y^2 \frac{S(f,z) S(\bar f, \bar z)}{4} dz d\bar z
\ee
where
\be
\label{eq:SchwarzianAndT}
S(f, z) \equiv \frac{f'''(z) f'(z) - \frac{3}{2} (f''(z))^2}{ (f'(z))^2} = \frac{12}{c} T(z)
\ee
is the Schwarzian derivative.  Thus the diffeomorphism maps pure AdS$_3$ to a general vacuum-sector metric with a non-vanishing stress tensor. Applying this operator valued diffeomorphism to (\ref{eq:PureAdSLogOfBulkBoundary}), we obtain the vacuum sector bulk-boundary OPE block\footnote{Note that in deriving this equation, we cut off the divergent near boundary integral at a constant $y$ plane as oppose to the constant $y_w$ plane used in (\ref{eq:PureAdSLogOfBulkBoundary}). This shift results in the $(w'(z_1) \bar{w}'(\bar{z}_1))^{h}$ factor that is essential to reproduce the transformation property of a boundary Virasoro primary.} 
\be
\phi(y, z_2, \bar{z}_2) \CO(z_1, \bar{z}_1)|_{vac} &=& (w'(z_1) \bar{w}'(\bar{z}_1))^{h} \left( \frac{u_2}{u_2^2 + (w_2-w_1)(\bar{w}_2- \bar{w}_1))} \right)^{2h},
\label{eq:BBPtfmed}
\ee
 where $u_2, w_2, \bar{w}_2$ are $u,w,\bar{w}$ in (\ref{eq:GeneralVacDiffeomorphism}) evaluated at $(y,z_2, \bar{z}_2)$, and  $w_1, \bar{w}_1$ are evaluated at $(0, z_1, \bar{z}_1)$. This is the key formulation of the bulk-boundary OPE block that will be used in this paper. 
 
 To evaluate (\ref{eq:BBPtfmed}), we need to solve equation (\ref{eq:SchwarzianAndT}) and its anti-holomorphic equivalent for the functions $f(z)$ and $\bar f(\bar z)$, determining them as functionals of the stress tensor operators $T(z), \bar T(\bar z)$.  Then we can evaluate equation (\ref{eq:PureAdSLogOfBulkBoundary}) by expanding the coordinates $u, w, \bar w$ in terms of $f, \bar f$.  To carry out this procedure explicitly in $1/c$ perturbation theory, we write
\be
f(z) = z+\frac{1}{c} f_1(z) + \frac{1}{c^2} f_2(z) + \cdots
\ee
and then solve for the $f_n$ in terms of $T$ using equation (\ref{eq:SchwarzianAndT}).  The first two $f_n$ are determined by the differential equations
\be
\label{eq:FirstFewfEquations}
f_1'''(z) - 12 T(z)&=& 0
\nn \\
2 f_1{}^{(3)}(z) f_1'(z)+3 f_1''(z){}^2-2 f_2{}^{(3)}(z) &=& 0
\ee
so for example, the first equation simply leads to $f_1(z) = -6 \int^z_0 dz' (z-z')^2 T(z')$.  Once we solve for the $f_n$, we can expand (\ref{eq:BBPtfmed}) to find the bulk-boundary OPE block\footnote{We took the logarithm because it renders computations simpler and more transparent \cite{Fitzpatrick:2016mtp}, but one could easily deal with the full OPE block directly instead.  Taking the logarithm of an operator is not at all innocuous in general, but due to our choice of regulator it will not present any problems.} 
\be\label{eq:PhiOExpand}
\log \phi(y,0,0) \CO(z, \bar z)  =  
2h \log\left(\frac{y}{z\bar{z}+y^{2}}\right) 
+ \underbrace{\frac{h\left(z\bar{z}+y^{2}\right) f_{1}'(z) - 2\bar{z} f_{1}(z) }{c\left(z\bar{z}+y^{2}\right)}}_{ K_{T} } 
+ \cdots
\ee
where the ellipsis denotes both the conjugate anti-holomorphic $K_{\bar T}$ terms as well as the perturbation series at order $1/c^2$ and above.  The order $1/c$ terms $K_T$ and $K_{\bar T}$  are line-integrals of the stress tensors $T$ and $\bar T$ against specific kernels.  For example, by combining terms above we find that
\be
\label{eq:KT}
K_T = \frac{12 h}{c} \int_0^z dz' \frac{(y^2 + z' \bar z)(z - z')}{y^2 + z \bar z} T(z') 
\ee
and similarly for the anti-holomorphic $K_{\bar T}$.  In the limit $y \to 0$ we recover the kernels \cite{Fitzpatrick:2016mtp} for the standard `boundary-boundary'  $\CO(z) \CO(0)$ OPE block.

 At the next order we would obtain the new kernels $K_{TT}, K_{\bar T \bar T}$, and also the mixed kernel $K_{T \bar T}$ which are computed explicitly in appendix \ref{app:GravitationalComputationsHigherOrders}.   The results are
\be
\label{eq:KTTandKTTb}
K_{TT} &=& \frac{72 h}{c^2} \int_{0}^{z}dz'\int_{0}^{z'}dz''\frac{\left(z-z'\right)^{2}\left(y^{2}+\overline{z}z''\right)^{2}}{\left(z\bar{z}+y^{2}\right)^{2}}T\left(z'\right)T\left(z''\right)
\nn \\
K_{T \bar T} &=&  -\frac{72hy^{2}}{c^{2}\left(z\bar{z}+y^{2}\right)^{2}}\int_{0}^{z}dz'\left(z-z'\right)^{2}\int_{0}^{\overline{z}}d\overline{z}'\left(\overline{z}-\overline{z}'\right)^{2}T\left(z'\right)\overline{T}\left(\overline{z}'\right)
\ee
for the bulk-boundary OPE blocks.  Note that the first reduces to the expected $\CO(z) \CO(0)$ kernel (compare to equation 4.40 of \cite{Fitzpatrick:2016mtp}) at this order, while the $K_{T \bar T}$ kernel vanishes as $y \to 0$, again matching with the expectations for the boundary (where OPE blocks factorize into holomorphic $\times$ anti-holomorphic parts).  In the next subsection we will present an alternative derivation that makes this matching explicit to all orders in $1/c$.

\subsection{Connection with Chern-Simons Wilson Lines}
\label{sec:ConnectionCSWilson}

The sl$(2)$  Wilson line formulation in \cite{Fitzpatrick:2016mtp} (based on the earlier work \cite{Verlinde:1989ua})  of the standard OPE block takes the form
\be
\CO(z_2) \CO(z_1) \supset W(z_2, z_1) =  \left. P \left\{ e^{\int_{z_1}^{z_2} dz^\mu A_\mu^a(z) L_x^a} \right\} \frac{1}{x^{2h}} \right|_{x=0} .
\label{eq:PrevWL}
\ee
First, we will review the notation and some of the results from \cite{Fitzpatrick:2016mtp}, and then we will see how to generalize (\ref{eq:PrevWL}) to the expression (\ref{eq:BulkWL}) above.

In the Wilson line expression (\ref{eq:PrevWL}),  $P$ indicates `path-ordering', the $A_\mu$s are the sl$(2)$ gauge fields, and the $L_x^a$ are the corresponding generators.  The variable $x$ is an auxiliary coordinate introduced so that $L_x^a$ can be written in an infinite dimensional representation, 
\be 
L^1 \cong L_{-1} = \partial_x , \qquad L^0 \cong L_0 = x \partial_x + h , \qquad L^{-1} \cong L_1 = \frac{1}{2} x^2 \partial_x + h x.
\ee
Equation (\ref{eq:PrevWL}) is the holomorphic part of the OPE block, and a similar anti-holomorphic piece is present in the full block. The boundary condition  on $A_\mu$ that leads to Virasoro symmetry is
\be
A_z |_{y=0} = L^1 + \frac{12}{c} T(z) L^{-1}.
\ee
For  boundary operators $\CO$, we can push the Wilson line connecting $\CO(z_2)$ and $\CO(z_1)$ onto the boundary so that only the above behavior at $y=0$ is necessary.  When we move one of the $\CO$s into the bulk to position $(y,z_2, \bar{z}_2)$, we will first take the Wilson line to be along the boundary from $(0,z_1, \bar{z}_1)$ to $(0,z_2, \bar{z}_2)$, and then to go directly to the bulk point $(y,z_2, \bar{z}_2)$ along constant $(z_2, \bar{z}_2)$.  Making the gauge choice $A_y=0$, the second part of the Wilson line is trivial.

  In \cite{Fitzpatrick:2016mtp}, it was shown that the path-ordered term $P \left\{ e^{\int_{z_1}^{z_2} dz^\mu A_\mu^a(z) L_x^a} \right\} $ could equivalently be written as
\be
e^{\frac{12 h}{c} \int_{z_1}^{z_2} dz T(z) x_T(z)}
\ee
after promoting $x$ everywhere to an operator $x_T(z_1)$ that is defined as the (operator valued) solution to the differential equation
\be
- x_T'(z) = 1 + \frac{6 T(z)}{c} x_T^2(z), \qquad x_T(z_2) = 0. 
\label{eq:xTconstr}  
\ee
In other words,
\be
 W( z_2, z_1) =
\left( e^{\int_{z_1}^{z_2} dz \frac{12 T(z)}{c} x_T(z)}\frac{1}{x_T(z_1)^2} \right)^h.
 \ee

A key point was that $x_T$ is closely related to  the uniformizing coordinates $f_T$ defined through the Schwarzian in \ref{eq:SchwarzianAndT}.  In particular,
\be
\frac{1}{x_T(z)} \equiv  \frac{f''_T(z)}{2f'_T(z)} - \frac{f'_T(z)}{f_T(z) - f_T(z_f)} .
\ee
automatically satisfies the constraint (\ref{eq:xTconstr}).  

Now we are ready to derive (\ref{eq:BulkWL}).  The starting point will be our general philosophy that $\phi$ in a general background follows from $\phi$ in the AdS vacuum combined with the operator-valued transformation (\ref{eq:GeneralVacDiffeomorphism}). This results in the bulk-boundary OPE block for $\phi \CO$ given by (\ref{eq:BBPtfmed}).
Our goal will be to write (\ref{eq:BBPtfmed}) in terms of the Wilson line building blocks.  For concision, let us define the exponential
\be
E_T \equiv e^{\frac{6}{c} \int_{z_1}^{z_2} dz' T(z') x_T(z')} .
\ee
From the constraint equation (\ref{eq:xTconstr}), we have 
\be
\log E_T = - \int_{z_1}^{z_2} dz' \frac{1+ x_T'(z)}{x_T(z)}  = \log \left( \frac{2 (f_T'(z_2))^{\frac{1}{2}} (f_T'(z_1))^{\frac{3}{2}}}{2 (f'_T(z_i))^2 + (f_T(z_2)- f_T(z_1)) f_T''(z_1)} \right) .
\ee
Furthermore, we see that the OPE block to has the correct semiclassical limit \cite{Fitzpatrick:2016mtp}
\be
W(z_2, z_1) \cong E_T^2 \frac{1}{x_T^2(z_1)} = \frac{f'_T(z_2) f'_T(z_1)}{(f_T(z_2) -f_T(z_1))^2}.  
 \ee 
It is now a straightforward matter to compare  (\ref{eq:BBPtfmed}) to the RHS of
\be
\left. P \left\{ e^{ \int_{z_1}^{z_2} dz A_z+ \int_{\bar{z}_1}^{\bar{z}_2} d\bar{z} \bar{A}_{\bar{z}}} \right\} \left( \frac{y}{y^2 + x \bar{x} }\right)^\Delta \right|_{x=\bar{x}=0}
 \cong E_T^{\Delta} \bar{E}_T^{\Delta} \left( \frac{y}{y^2 + x_T \bar{x}_T} \right)^\Delta
\ee
expanded out in terms of their dependence on $f_T, \bar{f}_T$ and confirm that they agree.\footnote{To be systematic, one can just solve for $x_T$ and $E_T$ in terms of $f'(z_2), f''(z_2)$ and substitute.}  Thus the conclusion is that the methods of \ref{sec:GravityComputationPhiO} are entirely consistent with those from \cite{Fitzpatrick:2016mtp}, and all of the techniques from that paper apply equally well to the bulk-boundary OPE.  In particular, one can compute the integration kernels $K_{T\cdots \bar T \cdots}$ very efficiently to high orders using the $x_T$ variables \cite{Fitzpatrick:2016mtp}; this is a significant technical improvement compared to solving equations like (\ref{eq:FirstFewfEquations}) directly.

We can go further and obtain a simple form for the generalization of (\ref{eq:BulkWL}) to the case of spinning fields and operators as well.  We relegate the details of the derivation to appendix \ref{app:SpinningBulkWL} and simply quote the result here:
\begin{equation}
\< A_{\mu_1 \dots \mu_\ell}(y,z_2, \bar{z}_2)   \CO_{h,\bar{h}}(z_1, \bar{z}_1) \> =  P \left\{ e^{ \int_{z_1}^{z_2} dz A_z+ \int_{\bar{z}_1}^{\bar{z}_2} d\bar{z} \bar{A}_{\bar{z}}} \right\} t^{\mu'_1}_{\mu_1} \dots t^{\mu'_\ell}_{\mu_\ell} K_{\mu'_1, \dots, \mu'_\ell}(y,x,\bar{x}) .
\label{eq:spinningBulkWL}
\end{equation}
Here, $\CO_{h,\bar{h}}$ is a boundary field of weight $(h,\bar{h})$ and $A_{\mu_1, \dots, \mu_\ell}$ is a bulk field with $\ell = h - \bar{h} \ge 0$ (a similar expression holds for $\ell \le 0$).  The factor $K_{\mu_1, \dots, \mu_\ell}$ is the vacuum AdS bulk-boundary propagator that we describe in detail in appendix \ref{app:SpinningBulkWL}, and the tensor $t^\mu_\nu$ is a diagonal matrix of the form
\be 
t^y_y = 1 , \qquad t^z_z = 1+ \frac{6}{c} \frac{T(z_2) y^4}{\bar{x}^2} , \qquad t^{\bar{z}}_{\bar{z}} = 1+\frac{6}{c} \bar{T}(\bar{z}_2) \bar{x}^2 . 
\label{eq:spinningextrapiece}
\ee 
Although we have not pursued it directly in this paper, these results can be used to study the reconstruction of massive spinning fields in the bulk.

\subsection{Evaluating Vacuum Sector Correlators} 
\label{sec:TTbarColleratorViaKernel}

In this section we will use the bulk-boundary OPE block to compute correlators of $\phi \CO$ with products of local stress tensors.  These correlators repackage all of the information about the overlap of $\phi \CO$ with the Virasoro vacuum sector.

Since $\<\phi\CO\>$ is simply given by the first term in equation (\ref{eq:PhiOExpand}), ie $\<\phi(y,0,0)\CO(z,\bar z)\>=\left(\frac{y}{y^2+z \bar z}\right)^{2h}$,  the simplest non-trivial correlator is $\< \phi \CO T \>$. It can be computed using (\ref{eq:KT}), giving
\be 
\label{eq:PhiOT}
\frac{ \< \phi(y,0,0) \CO(z,\bar{z}) T(z_1) \>}{ \< \phi(y,0,0) \CO(z,\bar{z}) \>}  
=\<K_TT(z_1)\>&=&
\frac{12 h}{c}  \int_{0}^{z}   dz' \frac{2 (z - z')(y^2 + z' \bar z)}{y^2 + z \bar z}  \frac{c}{2(z'-z_1)^4}
\nn \\  &=&
 \frac{hz^2 }{ z_1^3
   \left( z_1 - z \right)^2 } \left( z_1+ \frac{2 y^2(z_1-z) }{y^2+z \bar z}   \right)
\ee
The computation is suggested pictorially in figure \ref{fig:BulkBoundaryOPEBlock}.  This result matches bulk gravitational perturbation theory using AdS$_3$ Feynman diagrams in our chosen gauge, as we show explicitly in appendix \ref{app:BulkWittenDiagramComputation}.  This is no surprise, as the definition in equation (\ref{eq:OPEBlockAsPathIntegral}) essentially reproduces gravitational perturbation theory in a first quantized language.  

Naively, one might expect that this is only the first term in an infinite perturbation series for this correlation function.  However, the higher order contributions need to be regulated in a way that is consistent with Virasoro symmetry and with the fixed dimension $2h$ for the scalar CFT operator $\CO$.  In the context of Chern-Simons Wilson lines, we proposed a prescription for regulating multi-$T$ correlators in Appendix C.2 of \cite{Fitzpatrick:2016mtp} that produces the correct Virasoro OPE blocks. In appendix \ref{subsec:FromClassicalBackgroundtoMultiTVacuumCorrelators}, we argue that this regulator can be derived from the generating function of multi-T correlators. Applying this same regulator for the bulk-boundary OPE block, we find that all higher order contributions to $\< \phi \CO T \>$ vanish. Thus we claim that equation (\ref{eq:PhiOT}) is the exact result for this correlation function.  
We will provide another argument that equation (\ref{eq:PhiOT}) is exact in section \ref{sec:SymmetryDefinitionPhi}.

We can also compute the correlators $\< \phi \CO T T \>$ and $\< \phi \CO T \bar T \>$.  We provide details of the computations in appendix \ref{app:ComputationsBBOPEBlock}.  The results are that
\begin{footnotesize}
\begin{align}
 & \frac{\left\langle \phi\left(y,0,0\right)\mathcal{O}\left(z,\overline{z}\right)T\left(z_{1}\right)T\left(z_{2}\right)\right\rangle }{\left\langle \phi\left(y,0,0\right)\mathcal{O}\left(z,\overline{z}\right)\right\rangle }\nn\\
= & \frac{c}{2\left(z_{1}-z_{2}\right)^{4}}+\frac{h^{2}z^{4}\left(z_{1}z\bar{z}+y^{2}\left(3z_{1}-2z\right)\right)\left(z_{2}z\bar{z}+y^{2}\left(3z_{2}-2z\right)\right)}{z_{1}^{3}z_{2}^{3}\left(z-z_{1}\right){}^{2}\left(z-z_{2}\right)^{2}\left(z\bar{z}+y^{2}\right)^{2}}\\
 & +\frac{2hz^{2}\left(y^{2}z\bar{z}z_{1}z_{2}\left(z\left(z_{1}+z_{2}\right)-4z_{1}z_{2}\right)-z^{2}\bar{z}^{2}z_{1}^{2}z_{2}^{2}+y^{4}\left(zz_{1}z_{2}\left(z_{1}+z_{2}\right)-3z_{1}^{2}z_{2}^{2}-z^{2}\left(z_{1}-z_{2}\right)^{2}\right)\right)}{\left(z-z_{1}\right)\left(z_{2}-z\right)z_{1}^{3}z_{2}^{3}\left(z_{2}-z_{1}\right){}^{2}\left(z\bar{z}+y^{2}\right)^{2}}\nn
\end{align}
\end{footnotesize}
and
\begin{small}
\begin{align}
 & \frac{\left\langle \phi\left(y,0,0\right)\mathcal{O}\left(z,\overline{z}\right)T\left(z_{1}\right)\overline{T}\left(\overline{w}_{1}\right)\right\rangle }{\left\langle \phi\left(y,0,0\right)\mathcal{O}\left(z,\overline{z}\right)\right\rangle }\\
= & \frac{h^{2}z^{2}\bar{z}^{2}\left(y^{2}\left(3\bar{w}_{1}-2\bar{z}\right)+\bar{w}_{1}z\bar{z}\right)\left(y^{2}\left(3z_{1}-2z\right)+z_{1}z\bar{z}\right)}{z_{1}^{3}\overline{w}_{1}^{3}\left(z_{1}-z\right){}^{2}\left(\overline{w}_{1}-\bar{z}\right)^{2}\left(z\bar{z}+y^{2}\right)^{2}}+\frac{2hy^{2}z^{3}\bar{z}^{3}}{z_{1}^{3}\bar{w}_{1}^{3}\left(z-z_{1}\right)\left(\bar{w}_{1}-\bar{z}\right)\left(z\bar{z}+y^{2}\right)^{2}}\nn
\end{align}
\end{small}
These reduce to the expected $\CO \CO$ correlators as $y \to 0$.  

We should also emphasize that in the semiclassical limit, where we include sources with dimensions $h_H \propto c$ as $c \to \infty$, the correlators of $\phi$ will take the correct form.  This follows automatically from the definition of the OPE block in equation (\ref{eq:OPEBlockAsPathIntegral}) and the form of the vacuum metric in equation (\ref{eq:MetricwithT}).  We can compute correlators in a BTZ black hole background when we include a heavy operators $\CO_H(\infty) \CO_H(0)$, which lead to $\frac{1}{c}  \< T(z) \> = \frac{h_H}{c} \frac{1}{z^2}$ in the semiclassical limit.  We hope to study these correlators at a non-perturbative level in the future.

\section{An Exact Algebraic Definition for the Proto-Field $\phi(X)$}
\label{sec:SymmetryDefinitionPhi}

Our regulated bulk-boundary OPE block computes vacuum sector correlators exactly, and this suggests that we can obtain an exact definition for the proto-field $\phi$ built from the Virasoro primary $\CO$.   Now we provide this definition in a simple algebraic form, which originates from symmetry considerations.  Our $\phi(y,0,0)$ will satisfy
 \begin{equation}\label{eq:DefinitionOfPhi}
	L_m\phi(y,0,0)|0\>=0,\quad \bar{L}_m\phi(y,0,0)|0\>=0,\qquad m\ge2.
\end{equation}
This follows from the fact that $\phi$ is a scalar and the bulk points $(y,0,0)$ are invariant under bulk Virasoro transformations generated by $L_{m}$ with $m\ge2$.  We explain this in detail in section \ref{sec:DefinitionOfPhi} and appendix \ref{app:BulkVirasoroTransforms}. 

In the following discussion, we will write $\phi\left(y,0,0\right)$ as an expansion in small $y$ or the boundary OPE expansion (BOE)\footnote{In the conventional BCFT case, the bulk theory is a CFT (see \cite{Liendo:2012hy} for a nice discussion).  An identical expansion also applies when studying non-gravitational QFTs in AdS \cite{Paulos:2016fap}, because boundary dilatations correspond to a bulk isometry. When the bulk theory is gravitational, one cannot use pure symmetry or OPE type arguments to prove that this expansion converges, but our results suggest that it can be determined exactly to all orders in $y$ after bulk gauge fixing. It seems reasonable to expect that the small $y$ expansion of $\phi$ would have a finite radius of convergence, since no terms like $\sim e^{-1/y}$ are allowed by scaling symmetry.  We also explain in appendix \ref{app:GlobalBOEBasics} that symmetry arguments dictate this global conformal BOE result \cite{Paulos:2016fap, Nakayama:2015mva} }
\begin{equation}
\phi\left(y,0,0\right)\left|0\right\rangle =\sum_{N=0}^{\infty}y^{2h+2N}\left|\phi\right\rangle _{N}
\end{equation}
 where $|\phi\>_N$ is a level $N$ Virasoro descendant of $\CO$ in both holomorphic and anti-holomorphic sectors, since we are defining the proto-field $\phi$ to be made of $\CO$ and its descendants.\footnote{More generally, a full bulk field would have terms like $y^{h'+\bar{h}'} | \CO_{h', \bar{h}'} \>$, where $\CO_{h', \bar{h}'}$ is not a descendant of $\CO$.}
Then the conditions (\ref{eq:DefinitionOfPhi}) for
$\phi\left(y,0,0\right)$ will be equivalent to saying that $\left|\phi\right\rangle _{N}$
satisfies the following `bulk primary' conditions:
\begin{equation}\label{eq:Conditions}
L_{m}\left|\phi\right\rangle _{N}=0,\quad\overline{L}_{m}\left|\phi\right\rangle _{N}=0,\qquad\text{for }m\ge2.
\end{equation}
That is, $\phi\left(y,0,0\right)$ will be a sum over these operators
$\phi_{N}$ of different levels. The $\left|\phi\right\rangle _{N}$ is, in a sense, as close as possible to being a primary itself while still living in the bulk (ie it is a primary that is not quasi-primary).  It is an eigenstate of $L_0$ and is annihilated by all higher generators except $L_1$. We will say more about the non-trivial action of $L_1$ in appendix \ref{app:BulkVirasoroTransforms}. 

In particular, the  conditions (\ref{eq:Conditions}) imply that at each level, $|\phi\>_N$ factorizes, and can be written in the following form
\begin{equation}
\left|\phi\right\rangle _{N}=\lambda_{N}\mathcal{L}_{-N}\overline{\mathcal{L}}_{-N}\mathcal{\left|O\right\rangle },\qquad\lambda_{N}=\frac{\left(-1\right)^{N}}{N!\left(2h\right)_{N}}.
\end{equation}
where $\mathcal{L}_{-N}$ (and $\overline{\mathcal{L}}_{-N}$) are
linear combinations of products of holomorphic (and anti-holmorphic)
Virasoro generators at level $N$. Note that, the holomorphic and anti-holomorphic
conditions above are independent, which means that $\overline{\mathcal{L}}_{-N}$
will just be $\mathcal{L}_{-N}$ with $L$ replaced by $\overline{L}$.

The  conditions (\ref{eq:Conditions}) will uniquely determine $|\phi\>_N$ (or $\mathcal{L_{-N}}$) up to an overall normalization (will be explained below). The overall normalization of $|\phi\>_N$ is fixed by
\begin{equation} \label{eq:Normalization}
L_{1}^{N}\overline{L}_{1}^{N}\left|\phi\right\rangle _{N}=\left(-1\right)^{N}N!\left(2h\right)_{N}\left|\mathcal{O}\right\rangle.
\end{equation}
This normalization condition is based on the requirement that we correctly reproduce the vacuum correlator $\< \phi \CO \>$,that is, $\< \phi(y,0,0) \CO(z,\bar z) \>=\<\phi^\text{global}\CO\>=\left(\frac{y}{y^2+z \bar z}\right)^{2h}$. $\phi^\text{global}$ here is the global bulk field in the HKLL reconstruction \cite{Hamilton:2005ju}, which we explain in \ref{app:GlobalBOEBasics} is equivalent to 
\begin{equation}\label{eq:PhiGlobalBOE}
\phi^{\text{global}}\left(y,0,0\right)\left|0\right\rangle =\sum_{N=0}^{\infty}y^{2h+2N}\lambda_{N}L_{-1}^{N}\overline{L}_{-1}^{N}\left|\mathcal{O}\right\rangle .
\end{equation}
So the requirement that $\< \phi \CO \>=\<\phi^\text{global}\CO\>$ implies that 
\be
\mathcal{L}_{-N}|\CO\>=L_{-1}^{N}|\CO\> + \left(\text{other quasi-primaries and their descendants}\right)
\label{eq:cLNForm}
\ee
where the terms in the parenthesis are all orthogonal to $\CO$ and its global descendants, and will not contribute when computing $\< \phi \CO \>$. They are then fixed by solving (\ref{eq:Conditions}). When acting on $|\phi\>_N$ with $L_{1}^N\bar{L}_{1}^N$, the terms in the parenthesis will be killed, that's why we have the normalization condition (\ref{eq:Normalization}).\footnote{Specifically, $L_1^N \bar{L}_1^N |\phi\>_N=L_1^N\bar{L}_1^N|\phi^\text{global}\>=\lambda_NL_1^N\bar{L}_1^NL_{-1}^N\bar{L}_{-1}^N|\CO\>=(-1)^N N!(2h)_N|\CO\>$.} It's also true that in the large $c$ limit, our $\phi$ will reduce to $\phi^\text{global}$, as will be shown in \ref{sec:SolutionQP} that the terms in the parenthesis are suppressed at large $c$.

Now let us explain why the conditions (\ref{eq:Conditions}) uniquely determine $\mathcal{L}_{-N}$. It is easy to see that they are equivalent to the equations
\begin{equation}
\label{eq:ConstraintEquation}
L_{m_{1}}\cdots L_{m_{i}}\left|\phi\right\rangle _{N}=0,\qquad\sum_{i}m_{i}=N
\end{equation}
(and similarly for the anti-holomorphic part) where $L_{m_{1}}\cdots L_{m_{i}}$ represents the set of all level $N$ products of Virasoro generators with at least one $L_{m_{i}}$ with $m_{i}$ \ensuremath{\ge} 2. That is, $L_{m_{1}}\cdots L_{m_{i}}$ does not include $L_{1}^{N}$ . These conditions say that when $L_{m_{1}}\cdots L_{m_{i}}$ decreases the level of $\left|\phi\right\rangle _{N}$ back to level zero, the result vanishes. There are $p\left(N\right)-1$ independent ways (because we exclude $L_{1}^{N}$ ) to lower $|\phi\>_N$ to level zero, and thus $\left|\phi\right\rangle _{N}$ must satisfy $p\left(N\right)-1$ constraint equations. Since all the level $N$ descendants of $\left|\mathcal{O}\right\rangle $ form a $p\left(N\right)$ dimensional space, the above condition will fix the bulk field up to an overall constant. So $\phi\left(y,0,0\right)$ will be uniquely fixed by the constraints (\ref{eq:Conditions}) and the normalization condition (\ref{eq:Normalization}).

In section \ref{sec:DefinitionOfPhi} we motivate the definition of $\phi$ using Virasoro symmetry and the fact that  $\phi$ is a bulk scalar field. We then solve these conditions in various cases in section \ref{sec:SolvingPhi}. In section \ref{sec:RecursionRelationPhiOTT}, we show that our definition of $\phi(y,0,0) $ leads to a powerful recursive algorithm to compute correlators of the form of equation (\ref{eq:phiOTTb}), extending standard recursion relations for correlators of stress tensors with local CFT$_2$ primary operators. The results exactly agree with those obtained from the bulk-boundary OPE block in section \ref{sec:GravitationalWilsonLines}. 

\subsection{Virasoro Transformations of $\phi(X)$}\label{sec:DefinitionOfPhi}

In this section we will derive (\ref{eq:DefinitionOfPhi}) using the fact that $\phi$ must transform as a bulk scalar. This means that under a coordinate transformation, $\phi(z,\bar{z},y) \to \phi(z^\prime,\bar{z}^\prime,y^\prime)$. 

We would like to understand the transformation of $\phi$ under the action of Virasoro, which is defined on the boundary by $(z, \bar z) \to (g(z),\bar{g}(\bar{z}))$. We will constructively demonstrate that there is a unique extension of an infinitesimal boundary Virasoro transformation preserving the Fefferman-Graham gauge. Infinitesimally, we have
\be
 \epsilon L_m (y,z,\bar{z},S,\bar{S}) \equiv \epsilon(\delta_m y,\delta_m z,\delta_m \bar{z},\delta_m S,\delta_m \bar{S}).
\ee
where $S, \bar S$ parameterizes the metric and are defined in (\ref{eq:SchwarzianAndT}).
Then the transformation of $\phi$ under an infinitesimal Virasoro generator $L_m$ is determined by its scalar property:
\be
L_m \phi(z,\bar{z},y) = (\delta_m y \partial_y +\delta_m z \partial + \delta_m \bar{z} \bar{\partial}  )\phi(z,\bar{z},y)
\label{eq:ScalarProperty}
\ee
This transformation rule is expected to hold within correlation functions.   

We work out the gauge preserving extension of $L_m$ in Appendix \ref{app:BulkVirasoroTransforms}, with the result
\be
\delta_{m}y&=&\frac{1}{2}(m+1)yz^{m}\\
\delta_{m}z&=&\frac{z^{m-1}\left(\left(m^{2}+m+z^{2}S(z)\right)\bar{S}\left(\bar{z}\right)y^{4}-4z^{2}\right)}{y^{4}S(z)\bar{S}\left(\bar{z}\right)-4}\\
\delta_{m}\bar{z}&=&\frac{2m(m+1)y^{2}z^{m-1}}{y^{4}S(z)\bar{S}\left(\bar{z}\right)-4}\label{eq:LmTransformations}
\ee
We have verified that these results agree with the action of $L_{m}$ computed using contour integrals  \cite{Ginsparg} of the stress tensor correlators from section \ref{sec:TTbarColleratorViaKernel}. These results have several notable features. First, they reduce to the expected form of a Virasoro transformation on the boundary:
\be
\lim_{y\rightarrow 0}(\delta_m y,\delta_m z,\delta_m\bar{z})=(0,z^{m+1},0).
\ee
Secondly, the transformation on the coordinates depends on the starting metric through $(S,\bar{S})$. This fact is easy to understand because if no such dependency existed, then we would not be able preserve the Fefferman-Graham form of the metric in general.

The central feature of these transformations is that for $m\ge2$, points on the line $(y,0,0)$ are left invariant:
\be
\delta_m(y,0,0)=0\quad \text{for }m\ge2.
\ee
Using the scalar property (\ref{eq:ScalarProperty}), we find that
\be
L_m \phi(y,0,0)|0\rangle=0,\quad \text{for } m\ge2.
\label{eq:PrimaryDef}
\ee
Including the constraints from $\bar{L}_{\bar{m}}$, we arrive at conditions (\ref{eq:DefinitionOfPhi}) satisfied by  $\phi(y,0,0)$.

One can also motivate the conditions  (\ref{eq:DefinitionOfPhi}) satisfied by $\phi(y,0,0)$ by consideration of causality \cite{Kabat:2011rz, Kabat:2012av, Kabat:2013wga, Kabat:2016zzr}.  Correlators of $\phi(y,0,  0)$ with boundary stress tensors $T(z)$ necessarily have singularities of the form $\frac{1}{z^2}$, as the stress tensor must be sensitive to the energy-momentum `charge' of the bulk field, as well as $\frac{1}{z^3}$ singularities, since special conformal transformations move $\phi$ around in the bulk.\footnote{These singularities could move to a different location in a different gauge, but they cannot be eliminated entirely \cite{Kabat:2012av}.}  However, one may wish to forbid branch cuts and higher order singularities such as $\frac{1}{z^n}$ with $n \geq 4$. Our $\phi(y,0,0)$ is constructed to satisfy these requirements. The  conditions on $\phi$ are equivalent to stipulating that the singular terms in the OPE of the stress energy tensor $T(z)$ with $\phi(y,0,0)$ are 
\be
T\left(z\right)\phi\left(y,0,0\right)\sim\frac{L_{-1}\phi\left(y,0,0\right)}{z}+\frac{L_{0}\phi\left(y,0,0\right)}{z^{2}}+\frac{L_{1}\phi\left(y,0,0\right)}{z^{3}}.
\ee
So there will be no higher order singularities in correlators of $\phi$ with any number of  $T$. This property also holds for the individual components $\phi_N$.  One can also see this explicitly in the correlators $\<\phi\CO T\>$, $\<\phi\CO T T\>$, and $\<\phi\CO T \bar T\>$ that we computed using bulk-boundary OPE blocks in section \ref{sec:TTbarColleratorViaKernel}, where there are no singularities beyond $\frac{1}{z^3}$, including in the expansions of these expressions in $y$.

\subsection{Solving for $\phi(X)$ Explicitly}
\label{sec:SolvingPhi}

In this section, we will solve  the  conditions  (\ref{eq:Conditions}) and the normalization condition (\ref{eq:Normalization}) for $\phi(y,0,0)$ explicitly.  We will focus on the holomorphic part of $
|\phi\>_N=\lambda_N \mathcal{L}_{-N}\bar{\mathcal{L}}_{-N}|\CO\> $ and  solve for $\mathcal{L}_{-N}$, since $\bar{\mathcal{L}}_{-N}$ is just the anti-holomorphic conjugate.  In terms of $\mathcal{L}_{-N}$, the conditions are 
\begin{align}
L_m\mathcal{L}_{-N}|\CO\>&=0, \quad \quad \text{for } 2\le m\le N\\
L_1^N\mathcal{L}_{-N}|\CO\>&=N!(2h)_N|\CO\>
\end{align}
 We first provide an example at low orders in section \ref{sec:SolveLowOrders} , and then we obtain an exact, all orders solution in terms of orthogonal quasi-primaries in \ref{sec:SolutionQP}. We also solve these conditions in the large $c$ limit up to order $\CO(c^{-2})$ in appendix \ref{app:SolutionAtLargec}.

\subsubsection{Explicit Solutions at Low Orders}
\label{sec:SolveLowOrders}

It is obvious that $|\phi\>_0=|\CO\>$ and $|\phi\>_1=-\frac{1}{2h}L_{-1}\bar{L}_{-1}|\CO\>$, and so the first non-trivial case arises at the next level.  At level 2, an arbitrary $\mathcal{L}_{-2}$ is given by $\mathcal{L}_{-2}=b_1 L_{-1}^2+b_2 L_{-2}$ and
 the  conditions are 
\be
L_{2}\left(b_{1}L_{-1}^{2}+b_{2}L_{-2}\right)\left|\mathcal{O}\right\rangle & =&0,\\
L_{1}^2\left(b_{1}L_{-1}^{2}+b_{2}L_{-2}\right)\left|\mathcal{O}\right\rangle & =& 2!(2h)_2\left|\mathcal{O}\right\rangle.
\ee
  Solving these two equations for $b_1$ and $b_2$, we find
 \be\label{eq:MathematicalLMinus2}
\mathcal{L}_{-2}=\frac{(2 h+1) (c+8 h)}{\left(2h+1\right)c+2 h (8 h-5)}\left(L_{-1}^2-\frac{12h}{c+8h}L_{-2}\right)
 \ee
and $|\phi\>_2$ is given by $|\phi\>_2=\lambda_2 \mathcal{L}_{-2}\bar{\mathcal{L}}_{-2}|\CO\>$. One can continue this process at higher orders (we also computed $|\phi\>_3$ and $|\phi\>_4$ in Appendix \ref{app:Level3Level4Solution}.), although the explicit expressions become rather complicated.  Instead we will see how to solve these equations in general in terms of quasi-primaries.

\subsubsection{Solution in Terms of Quasi-Primaries}
\label{sec:SolutionQP}

We know that $\left|\phi\right\rangle _{N}$ can be written as the sum of the level $N$ descendants of $\mathcal{O}$.  These descendants can be decomposed into quasi-primaries (global primaries) and their global conformal descendants.  In this subsection, we will show that the coefficients in this decomposition are determined by the norms of the quasi-primaries. We already saw an obvious example in the global case, as the global descendant $L_{-1}^N\bar{L}_{-1}^N\left|\mathcal{O}\right\rangle $ appears as
\be
\left|\phi\right\rangle _{N}\supset\left(-1\right)^{N}\frac{1}{N!\left(2h\right)_{N}}L_{-1}^{N} \bar L_{-1}^N \left|\mathcal{O}\right\rangle =\left(-1\right)^{N}\frac{L_{-1}^{N} \bar L_{-1}^N \left|\mathcal{O}\right\rangle }{\left|L_{-1}^{N}\mathcal{O}\right|^2}
\ee
where $\left|L_{-1}^{N}\mathcal{O}\right|^2 \equiv\left\langle \mathcal{O}|L_{1}^{N}L_{-1}^{N}|\mathcal{O}\right\rangle = N!(2h)_N .$
We will show that phenomenon is a general feature of the quasi-primary decomposition.

Suppose $\mathcal{L}_{-N}^{\text{quasi}}$ is a linear combination of Virasoro generators that acts on $\left|\mathcal{O}\right\rangle $
to create a quasi-primary at level $N$, with the coefficient of $L_{-1}^{N}$ in $\mathcal{L}_{-N}^{\text{quasi}}$ normalized to $1$. For example, at level two there is a unique $\mathcal{L}_{-2}^{\text{quasi}}=L_{-1}^{2}-\frac{2\left(2h+1\right)}{3}L_{-2}$. Since there are many quasi-primaries\footnote{The number of quasi-primaries at level $N$ is $p\left(N\right)-p\left(N-1\right)$, where $p\left(N\right)$ is the number of partitions of $N$.}  at level $N$, we will take the quasi-primary created by our chosen generator $\mathcal{L}_{-N}^{\text{quasi}}$ to be orthogonal to all of the other level $N$ quasi-primaries, and normalized to contain exactly $L_{-1}^N$. 

In what follows we will treat the holomorphic and anti-holomorphic descendants of $\CO$ separately, since at each level $\phi_N$ factorizes.  Then we will combine the holomorphic and anti-holomorphic pieces and correctly normalize them.
Let us define the coefficient of $\mathcal{L}_{-N}^{\text{quasi}}\left|\mathcal{O}\right\rangle $ in $\left|\phi\right\rangle _{N}=\lambda_N\mathcal{L}_{-N}|\CO\>$\footnote{Via an abuse of notation, here $\left|\phi\right\rangle _{N}=\lambda_N\mathcal{L}_{-N}|\CO\>$, but it should be clear from the context  whether $\bar{\mathcal{L}}_{-N}$ is included in the definition of $|\phi\>_N$ or not.} to be $b_{N}$, that is 
\be
\left|\phi\right\rangle _{N}\supset b_{N}\mathcal{L}_{-N}^{\text{quasi}}\left|\mathcal{O}\right\rangle 
\ee
When we take the inner product of $\left|\phi\right\rangle _{N}$ with $\mathcal{L}_{-N}^{\text{quasi}}\left|\mathcal{O}\right\rangle $,
we obtain 
\be
\left\langle \mathcal{O}\right|\left(\mathcal{L}_{-N}^{\text{quasi}}\right)^{\dagger}\left|\phi\right\rangle _{N} & =b_{N}\left\langle \mathcal{O}\left|\left(\mathcal{L}_{-N}^{\text{quasi}}\right)^{\dagger}\mathcal{L}_{-N}^{\text{quasi}}\right|\mathcal{O}\right\rangle \equiv b_{N}\left|\mathcal{L}_{-N}^{\text{quasi}}\mathcal{O}\right|^2,
\ee
because $\mathcal{L}_{-N}^{\text{quasi}}\left|\mathcal{O}\right\rangle $
is orthogonal to all other states in $\left|\phi\right\rangle _{N}$. 

Now, using the conditions defining $\phi_N$, we have 

\begin{align}
\left\langle \mathcal{O}\right| \left( \left(\mathcal{L}_{-N}^{\text{quasi}}\right)^{\dagger}-L_{1}^{N} \right) \left|\phi\right\rangle _{N} & =0\label{eq:QuasiKillPhi}
\end{align}
because all of the terms in $\left(\mathcal{L}_{-N}^{\text{quasi}}\right)^{\dagger}-L_{1}^{N}$
will include at least one $L_{m}$, with $m\ge2$, and according to the conditions (\ref{eq:ConstraintEquation}), these terms will all annihilate $\left|\phi\right\rangle _{N}$.
Using the normalization condition
\be
L_{1}^{N}\left|\phi\right\rangle _{N}=\frac{\left(-1\right)^{N}}{N!\left(2h\right)_{N}}L_{1}^{N}L_{-1}^{N}\left|\mathcal{O}\right\rangle =\left(-1\right)^{N}\left|\mathcal{O}\right\rangle ,
\ee
equation (\ref{eq:QuasiKillPhi})  leads to
\be
b_{N}=\frac{(-1)^N}{\left|\mathcal{L}_{-N}^{\text{quasi}}\mathcal{O}\right|^2}.
\ee
So we have shown that the coefficient of the level $N$ quasi-primary $\mathcal{L}_{-N}^{\text{quasi}}$ will be given by the inverse of its norm. Actually, one can show that this is also true even for the global descendants of the quasi-primaries. The holomorphic part of $\left|\phi\right\rangle _{N}$ will be given in the following form:\footnote{It is easy to see $\left|L_{-1}^{m}\mathcal{L}_{-\left(N-m\right)}^{\text{quasi}}\mathcal{O}\right|^2=m!(2(h+N-m))_{m}\left|\mathcal{L}_{-\left(N-m\right)}^{\text{quasi}}\mathcal{O}\right|^2$.}
\begin{footnotesize}
\be
|\phi\>_N \propto (-1)^N\left(\frac{L_{-1}^{N}}{\left|L_{-1}^{N}\mathcal{O}\right|^2}+\frac{\mathcal{L}_{-N}^{\text{quasi}}}{\left|\mathcal{L}_{-N}^{\text{quasi}}\mathcal{O}\right|^2}+\frac{L_{-1}\mathcal{L}_{-\left(N-1\right)}^{\text{quasi}}}{\left|L_{-1}\mathcal{L}_{-\left(N-1\right)}^{\text{quasi}}\mathcal{O}\right|^2}+\cdots+\frac{L_{-1}^{m}\mathcal{L}_{-\left(N-m\right)}^{\text{quasi}}}{\left|L_{-1}^{m}\mathcal{L}_{-\left(N-m\right)}^{\text{quasi}}\mathcal{O}\right|^2}+\cdots\right)|\mathcal{O}\>.
\nn
\ee
\end{footnotesize}
Including the anti-holomorphic part and accounting for the overall coefficient (ie requiring the coefficient of $L_{-1}^N\bar{L}_{-1}^N$ to be $\lambda_N=\frac{(-1)^N}{\left|L_{-1}^N\CO\right|^2}$), we find
\begin{align}
\label{eq:SolutionforPhiQP}
|\phi\>_N & =\left(-1\right)^{N}\left|L_{-1}^{N} \mathcal{O}\right|^2  \left(\frac{L_{-1}^{N}}{\left|L_{-1}^{N}\mathcal{O}\right|^2 }+\frac{\mathcal{L}_{-N}^{\text{quasi}}}{\left|\mathcal{L}_{-N}^{\text{quasi}}\mathcal{O}\right|^2}+\frac{L_{-1} \mathcal{L}_{-\left(N-1\right)}^{\text{quasi}}}{\left|L_{-1}\mathcal{L}_{-\left(N-1\right)}^{\text{quasi}}\mathcal{O}\right|^2}+\cdots\right) 
\nn \\
 & \quad  \times \left(\frac{ \bar L_{-1}^{N}}{\left| \bar L_{-1}^{N}\mathcal{O}\right|^2 }+\frac{\bar \CL_{-N}^{\text{quasi}}}{\left|\bar \CL_{-N}^{\text{quasi}}\mathcal{O}\right|^2}+\frac{\bar L_{-1} \bar \CL_{-\left(N-1\right)}^{\text{quasi}}}{\left| \bar L_{-1} \bar \CL_{-\left(N-1\right)}^{\text{quasi}}\mathcal{O}\right|^2}+\cdots\right) |\CO\>
\end{align}
as the exact solution for $|\phi\>_N$ in terms of orthogonal quasi-primaries with our chosen normalization.  Note that in a large $c$ expansion, the norms of the non-trivial quasi-primaries (and their descendants) will be proportional to positive powers of $c$, so that their contributions will be suppressed.  But at finite $c$, or for $h \gtrsim c$, their contributions will be on equal footing with the global conformal descendants $\phi$.

As an illustration of the above result, $\lambda_2\mathcal{L}_{-2}$ in $|\phi\>_2$ derived in equation (\ref{eq:MathematicalLMinus2}) of last section can be written in the following form:
\begin{align}
\lambda_{2}\mathcal{L}_{-2} & =\frac{L_{-1}^{2}}{2!\left(2h\right)_{2}}+\frac{L_{-1}^{2}-\frac{2\left(2h+1\right)}{3}L_{-2}}{\frac{2}{9}\left(2h+1\right)\left(c\left(2h+1\right)+2h(8h-5)\right)} =\frac{L_{-1}^{2}}{\left|L_{-1}^{2}\mathcal{O}\right|^{2}}+\frac{\mathcal{L}_{-2}^{\text{quasi}}}{\left|\mathcal{L}_{-2}^{\text{quasi}}\right|^{2}}
\end{align}
with $|\mathcal{L}^{\text{quasi}}_{-2}\CO|^2=\frac{2}{9} (2 h+1) \left((2h+1)c+2 h (8 h-5)\right)$. We also explicitly compute $|\phi\>_3$ and $|\phi\>_4$ in Appendix \ref{app:Level3Level4Solution}.

\subsection{Recursion Relation for Stress-Tensor Correlators}
\label{sec:RecursionRelationPhiOTT}

In section (\ref{sec:TTbarColleratorViaKernel}) we computed correlators of the form $\left\langle \phi\mathcal{O}T\cdots\overline{T}\cdots\right\rangle$ using the bulk-boundary OPE block. In this section, we will derive a recursion relation that can be used to calculate these correlators.  Specifically, we will express correlators with $n+1$ stress tensors in terms of a differential operator acting on correlators with fewer stress tensors.  This relation generalizes the well-known case of $\left\langle \mathcal{O}\mathcal{O}T\cdots\overline{T}\cdots\right\rangle$ correlators \cite{DiFrancesco:1997nk}, which can be derived recursively from the two point function $\left\langle \mathcal{O}\mathcal{O}\right\rangle$ using the Virasoro Ward identity. 

Suppose we know the correlator with $n$ insertions of $T$ and $m$ insertions of $\overline{T}$, 
\be
G_{n,m}\equiv\left\langle T\left(z_{1}\right)\cdots T\left(z_{n}\right)\overline{T}\left(\overline{w}_{1}\right)\cdots\overline{T}\left(\overline{w}_{m}\right)\mathcal{O}\left(z,\overline{z}\right)\phi\left(y,0,0\right)\right\rangle ,
\ee
and now we consider the case of one more $T$ insertion, 
\be
G_{n+1,m}\equiv\left\langle T\left(z_{1}\right)\cdots T\left(z_{n}\right)T\left(z_{n+1}\right)\overline{T}\left(\overline{w}_{1}\right)\cdots\overline{T}\left(\overline{w}_{m}\right)\mathcal{O}\left(z,\overline{z}\right)\phi\left(y,0,0\right)\right\rangle .
\ee
A key feature of stress tensor correlators such as $G_{n+1,m}$ is that as $z_{n+1} \to \infty$, the correlator vanishes.  This means that $G_{n+1,m}$ is completely determined by its poles in the $z_{n+1}$ variable.  Thus $G_{n+1,m}$ can be computed by taking the OPE of $T\left(z_{n+1}\right)$ with all the other operators in $G_{n+1,m}$ and only keeping the singular terms. We know the singular terms in the OPE of $T\left(z_{n+1}\right)$ with $\mathcal{O}\left(z,\overline{z}\right)$ and $T\left(z_{i}\right)$, which are 
\begin{align*}
T\left(z_{n+1}\right)\mathcal{O}\left(z,\overline{z}\right) & \sim\frac{h\mathcal{O}\left(z,\overline{z}\right)}{\left(z_{n+1}-z\right)^{2}}+\frac{\partial_{z}\mathcal{O}\left(z,\overline{z}\right)}{z_{n+1}-z},\\
T\left(z_{n+1}\right)T\left(z_{i}\right) & \sim\frac{c}{2\left(z_{n+1}-z_{i}\right)^{4}}+\frac{2T\left(z_{i}\right)}{\left(z_{n+1}-z_{i}\right)^{2}}+\frac{\partial T\left(z_{i}\right)}{z_{n+1}-z_{i}}.
\end{align*}
The conditions of equation (\ref{eq:Conditions}) tell us that the singular terms in the OPE of $T\left(z_{n+1}\right)$ with $\phi\left(y,0,0\right)$ are given by 
\be
T\left(z_{n+1}\right)\phi\left(y,0,0\right)\sim\frac{L_{1}\phi\left(y,0,0\right)}{z_{n+1}^{3}}+\frac{L_{0}\phi\left(y,0,0\right)}{z_{n+1}^{2}}+\frac{L_{-1}\phi\left(y,0,0\right)}{z_{n+1}}.
\ee
 Writing $\left|\phi\right\rangle $ as a sum over $\left|\phi\right\rangle _{N}$,
that is $\left|\phi\right\rangle =\sum_{N=0}^{\infty}y^{2h+2N}\left|\phi\right\rangle _{N}$,
we know that the effect of $L_{0}$ on $\left|\phi\right\rangle $
is to pull down a factor of $h+N$ for each $\left|\phi\right\rangle _{N}$.
This is equivalent to taking the derivative with respect to $y$,
so we have 
\be
L_{0}\phi\left(y,0,0\right)=\frac{1}{2}y\partial_{y}\phi\left(y,0,0\right).
\ee
And it's easy to see that 
\be
L_{-1}\phi\left(y,0,0\right)=\partial_{x}\phi\left(y,x,\overline{x}\right)|_{x,\overline{x}=0}.
\ee
Because of translation invariance, the action of $L_{-1}$ on $\phi\left(y,0,0\right)$ is equal to a holomorphic partial derivative of all of the other operators, namely $\mathcal{O}\left(z,\overline{z}\right)$ and other $T$s in the correlator $G_{n,m}$. 

The term $L_{1}\phi\left(y,0,0\right)$ is more subtle.  In general, at finite $c$ we cannot write it as a simple differential operator acting on $\phi\left(y,0,0\right)$ itself (see appendix \ref{app:BulkVirasoroTransforms} for more details). But since $L_{-1}$ annihilates the vacuum, ie $\left\langle 0\right|L_{1}=\left(L_{-1}\left|0\right\rangle \right)^{\dagger}$
= 0, we can commute $L_{1}$ with all the other operators on the left. Since
we know the action of $L_{1}$ on $\mathcal{O}$ and the stress tensor,\footnote{The commutators of $L_{1}$ with $\mathcal{O}$ and $T$ are simply
\begin{align*}
\left[L_{1},\mathcal{O}\left(z,\overline{z}\right)\right] & =z\left(2h+z\partial_{z}\right)\mathcal{O}\left(z,\overline{z}\right),\\
\left[L_{1},T\left(z_{i}\right)\right] & =z_{i}\left(4+z_{i}\partial_{z_{i}}\right)T\left(z_{i}\right).
\end{align*}
} we can evaluate its action on $\phi$ within the vacuum sector correlator
$G_{n,m}$.

Combining all the above facts, we obtain a recursion relation for computing $G_{n+1,m}$ from $G_{n,m}$ and $G_{n-1,m}$:
\begin{footnotesize}
\begin{align}
\mathcal{}G_{n+1,m}= & \left(-\frac{\partial_{z}+\sum_{i=1}^{n}\partial_{z_{i}}}{z_{n+1}}+\frac{\frac{y}{2}\partial_{y}}{z_{n+1}^{2}}-\frac{z\left(2h+z\partial_{z}\right)}{z_{n+1}^{3}}+\sum_{i=1}^{n}\frac{-z_{i}\left(4+z_{i}\partial_{z_{i}}\right)}{z_{n+1}^{3}}\right)G_{n,m}
\nn \\
 & +\left(\frac{h}{\left(z_{n+1}-z\right)^{2}}+\frac{\partial_{z}}{\left(z_{n+1}-z\right)}+\sum_{i=1}^{n}\left(\frac{2}{\left(z_{n+1}-z_{i}\right)^{2}}+\frac{\partial_{z_{i}}}{z_{n+1}-z_{i}}\right)\right)G_{n,m}\\
 & + \sum_{i=1}^{n}\frac{\left\langle T\left(z_{1}\right)T\left(z_{2}\right)\cdots T\left(z_{i-1}\right)T\left(z_{i+1}\right)\cdots T\left(z_{n}\right)\overline{T}\left(\overline{w}_{1}\right)\cdots\overline{T}\left(\overline{w}_{m}\right)\mathcal{O}\left(z,\overline{z}\right)\phi\left(y,0,0\right)\right\rangle }{2\left(z_{n+1}-z_{i}\right)^{4}}
\nn
\end{align}
\end{footnotesize}
We display the origin of all of these terms in appendix \ref{app:underbraceRecursion}.  In appendix \ref{app:underbraceRecursion}, we also use this recursion relation to easily reproduce the correlators $\<\phi\CO T\>$, $\<\phi\CO TT\>$ and $\<\phi\CO T\bar{T}\>$ computed in section \ref{sec:TTbarColleratorViaKernel} using the bulk-boundary OPE block.

One can derive an identical recursion relation with $T \leftrightarrow \bar T$ for adding insertions of the anti-holomorphic stress tensor.  Together, these relations precisely determine all vacuum sector correlators of $\phi \CO$.  In other words, one can view these recursion relations as an alternative definition for the proto-field $\phi$, which is entirely equivalent to the definition (\ref{eq:PrimaryDef}) and the bulk-boundary OPE block prescription and accompanying regulator from section \ref{sec:GravitationalWilsonLines}. 
\section{Discussion}

It is natural to conjecture \cite{Kabat:2011rz} that complete, interacting scalar fields $\Phi(X)$ in AdS$_3$ should be written as
\be
\Phi(X) = \sum_\CO \lambda_\CO \phi_\CO(X)
\ee
where the sum runs over all scalar CFT$_2$ primaries, and the coefficients $\lambda_\CO$ are constrained by consistency and causality  \cite{Kabat:2011rz, Kabat:2012av, Kabat:2013wga, Kabat:2016zzr}.  Our work does not shed much light on the questions of existence, (non-)uniqueness, and efficient determination of the $\lambda_\CO$.    

However, we have proposed a formula for the local AdS$_3$ proto-field operator $\phi_\CO$ built from a specific CFT$_2$ primary $\CO$ and its Virasoro descendants.\footnote{This may be enough to reconstruct the (toy?) case of a CFT with a low-energy spectrum that is dual to AdS$_3$ gravity coupled to a free bulk scalar field.}  We argued that our choice of $\phi_\CO$ has a number of desirable properties, including healthy vacuum-sector correlators that match bulk Witten diagrams, a natural interpretation in any semiclassical vacuum geometry, and Virasoro symmetry transformations implemented as bulk diffeomorphisms.    But perhaps the most surprising and intriguing aspect of our analysis is that we have determined $\phi_\CO$ exactly, based on the extremely simple condition of equation (\ref{eq:DefinitionOfPhi}).  

Profound lore based on diffeomorphism gauge redundancy and black hole physics suggests that local observables in gravitational theories may be ambiguous\footnote{For an example of an interesting recent discussion see \cite{Jafferis:2017tiu}.} or ill-defined.  Hopefully our formalism will provide a context where these ideas can be made more precise.  It may be that AdS$_3$ differs significantly from the case of higher dimensions (or AdS$_3 \times X$ spacetimes), where most aspects of bulk gravitational physics cannot be fixed by symmetry, and the gravitational dynamics can depend on many parameters.  In CFT$_{\geq 3}$ this difference arises because the OPE of the stress tensor is largely unconstrained, in marked contrast with the CFT$_2$ case.

\subsection*{How Non-Local is $\phi$?}

Our construction of $\phi$ was based on a series expansion in the radial coordinate $y$, which may be viewed as a gravitational version of the boundary operator expansion of boundary CFT.  The non-locality of $\phi$ (as a CFT operator) arises from the fact that it has been expressed as an infinite sum of local operators.  In the global conformal case, one can precisely relate the standard HKLL smearing function to the boundary operator expansion (see appendix \ref{app:GlobalBOEBasics}), making the non-locality of $\phi$ manifest. 
The extent of the non-locality displayed by the exact Virasoro $\phi$  remains less clear.  It should be possible to evaluate this region by computing correlators of $\phi$ with local CFT operators and investigating the convergence properties of the infinite sum.\footnote{This suggests an amusing exercise -- one might Borel resum the boundary operator expansion for $\phi$.  It seems plausible that the summation defining the Borel series (operator) would appear local, in the sense that its series expansion would converge in correlators with all other local operators. }   There may be a more direct method involving a non-perturbative generalization of the smearing procedure.  

These questions will be of particular interest when we move from Euclidean to Lorentzian signature.  Lorentzian CFT correlators can be obtained from their Euclidean counterparts by analytic continuation, but we do not know to what extent this holds for bulk dynamics.  At the very least we will need to have a better understanding of bulk diffeomorphisms, including large transformations to new gauges.  From the bulk or Wheeler-DeWitt perspective, the formation and evaporation of a black hole can be pure gauge!  

Many recent works have focused on the relationship between bulk and boundary domains of dependence \cite{Czech:2012bh, Bousso:2012mh, Morrison:2014jha, Dong:2016eik, Faulkner:2017vdd} in Lorentzian signature.  Some of this work \cite{Almheiri:2014lwa} was motivated by putative ambiguities in bulk reconstruction associated with the fact that a bulk operator $\phi(X)$ can be expressed using smearing functions supported on different boundary domains \cite{Hamilton:2005ju}.  These ambiguities do not exist for non-gravitational AdS field theory and its non-local boundary dual, as in this case $\phi(X)$ is precisely well-defined.    Thus it appears that these ambiguities can only arise from non-perturbative gravitational effects.  It would be interesting to exhibit such effects explicitly and to characterize their physical significance in the bulk; perhaps this is possible in AdS$_3$/CFT$_2$.

\subsection*{Bulk Locality and Horizons}

The primary motivation for studying $\phi$ is to investigate bulk locality and physics near and beyond black hole horizons \cite{Mathur:2009hf, Mathur:2010kx, Almheiri:2012rt, Almheiri:2013hfa}.

The breakdown of bulk locality can be analyzed using scattering in AdS/CFT \cite{GGP, Maldacena:2015iua}.  However, one can attack the problem much more directly by studying the operator product $\phi(X) \phi(Y)$ and its expectation value.  The correlator $\< \phi(X) \phi(Y) \>$ can differ greatly from that of a free bulk scalar field because it includes the exchange of arbitrary Virasoro descendants of $\CO$, or in the language of multi-trace operators, states such as `$[T\partial^2 T \bar T  \bar \partial^4 \CO]$' built from the OPE of $\CO$ with any number of stress tensors.  Since the contribution of these states has been fixed exactly, one can compute $\< \phi(X) \phi(Y) \>$ at finite operator dimension $h$ and central charge $c$ and as a function of the geodesic separation between the bulk operators.  When $h \ll c$ one might hope to see the breakdown of bulk locality at the Planck scale, and for heavy operators with $h \gg c$ one might see indications of the horizon radius (or some other pathology associated with bulk fields dual to very heavy CFT states).  More generally, we would expect that the bulk OPE expansion of $\phi(X) \phi(Y)$ does not exist.  

We can also use correlators like $\< \phi \CO_H \CO_H \>$ and $\< \phi \phi \CO_H \CO_H \>$  to probe the vicinity of black hole horizons.  In these and other high-energy states, we may find that $\phi$ breaks down deep in the bulk, and it will be interesting to understand when and how.  Previously it was unclear how to study such observables in a non-trivial way, since it seemed that one would need to rely on bulk perturbation theory to define them.  It appears that our construction surmounts this particular obstacle.  

Many aspects of black hole thermodynamics are encoded in the Virasoro algebra at large central charge \cite{Fitzpatrick:2014vua, Fitzpatrick:2015zha, Anous:2016kss, Asplund:2014coa, TakayanagiExcitedStates}, including various non-perturbative effects that resolve or ameliorate information loss problems \cite{Fitzpatrick:2016ive, Fitzpatrick:2016mjq, Chen:2017yze}.  This means that it should be possible to learn about bulk physics in the presence of black holes using Virasoro technology.  

Furthermore, general considerations  \cite{Balasubramanian:2007qv} borne out by non-perturbative investigations of Virasoro blocks \cite{Chen:2017yze} show that in Euclidean space,   pure  high-energy quantum states look very different from the BTZ black hole solution in the vicinity of the horizon.  This follows from the fact that thermal and BTZ correlators are periodic in Euclidean time, while pure state correlators display completely unsuppressed violations of this periodicity \cite{Chen:2017yze}.  Thus we have reason to believe that correlators like $\< \phi \phi \CO_H \CO_H \>$ will tell us about interesting structures near the Euclidean horizon.  By decomposing correlators into Virasoro blocks, we can learn which of these effects are universal, and which depend on the details of the CFT data.  

Of course the real question is whether black hole horizons appear innocuous to infalling \emph{Lorentzian} observers.  We hope to address some of these questions soon.

\section*{Acknowledgments} 
 
We would like to thank Tarek Anous,  Bartek Czech, Xi Dong, Ethan Dyer, Diego Hofman, Daniel Harlow, Tom Hartman, Daniel Jafferis, Shamit Kachru, David E. Kaplan, Zuhair Khandker, Nima Lashkari, Alex Maloney, Sam McCandlish, Joao Penedones, Suvrat Raju, Mukund Rangamani, Wei Song, Douglas Stanford, Matt Walters, and Junpu Wang for discussions, and Ibou Bah for discussions and comments on the draft.  We especially thank Miguel Paulos for discussing and sharing his unpublished notes with us.  We also thank the ICTP-SAIFR for hospitality while portions of this work were completed.  DL would like to thank the Aspen Center for Physics for support and hospitality while portions of this work were completed, which is supported by National Science Foundation grant PHY-1607611. ALF was supported in part by the US Department of Energy Office of Science under Award Number DE-SC-0010025.   JK, NA and HC  have been supported in part by NSF grant PHY-1454083.  ALF, JK and DL were also supported in part by the Simons Collaboration Grant on the Non-Perturbative Bootstrap.

\appendix

\section{Background and Review}

Here we collect fairly elementary results that may be of interest to some readers, and that provides some useful background material for the main body of the paper.  

\subsection{Klein-Gordon Equation from the Worldline Path Integral}
\label{app:KleinGordon}

Here we review  that first-quantized particles have propagators that satisfy the Klein-Gordon equation.  This follows implicitly from the equivalence between the two-point correlator of a free quantum field and the first-quantized propagator.  But we can also understand it more directly.

The first quantized propagator is
\be
K(x_f, x_i) = \int^{x_f}_{x_i} \CD x(t) e^{-m \int_i^f \sqrt{g_{\mu \nu} \dot x^\mu \dot x^\nu } }
\ee
Since $K$ propagates wavefunctions in time, it satisfies the Schrodinger equation, and the idea is that this equation is equivalent to the Klein-Gordon equation.  For this purpose we need to define a temporal direction for quantization, though we will find that this choice is irrelevant as the Klein-Gordon equation is covariant.  It's convenient to choose $t = \log y$ in our AdS case, so that we have a Lagrangian proportional to $\sqrt{g_{\mu \nu} \dot x^\mu \dot x^\nu } = \sqrt{1 + \dot x^i \dot x_i}$.    Then the canonical momenta are
\be
p^i = -\frac{m \dot x^i}{ \sqrt{1 + \dot x^i \dot x_i} }
\ee
and we find that the Hamiltonian is $H = \sqrt{m^2 - p^i p^j g_{ij}}$.  Interpreting the canonical momenta as covariant derivatives $p_i = \nabla_i$, the square of the Schrodinger equation $\partial_t^2 K = ( \nabla_i \nabla^i + m^2) K$ is the Klein-Gordon equation in our chosen coordinate system.  Note that one might try to identify $p_i = -i \partial_i$ as ordinary derivatives, but this leads to operator ordering ambiguities after quantization since $g_{ij}$ depends on $x^i$.  The choice $p_i = -i \nabla_i$ resolves these issues; equivalently, there is a particular choice of ordering of factors of $g_{ij}$ and $p_i \rightarrow -i \partial_i$ in the Hamiltonian that is equivalent to just setting $p_i = -i \nabla_i$.  Presumably, this choice should be correctly determined by a proper treatment of the path integral.  

\subsection{Geodesics in Euclidean AdS$_3$}

We would like to identify the geodesics in pure Euclidean AdS$_3$.  The analysis is most elegant using the embedding space coordinates
\be
\label{eq:EmbeddingGlobalPoincareEuclidean}
X_0 &=& R \frac{\cosh \tau}{\cos \rho} = \frac{1}{2} \left(\frac{ y^2 + z \bar z + R^2}{y} \right)  \\
X_{3} &=& R \frac{\sinh \tau}{\cos \rho} = \frac{1}{2} \left( \frac{y^2 + z \bar z  -  R^2}{y} \right)  \nn \\
X_{z}&=& R \tan \rho e^{i \theta} = \frac{R}{y} z \nn  \\
X_{\bar z} &=& R \tan \rho e^{-i \theta} = \frac{R}{y} \bar z \nn 
\ee
where we will set the AdS scale $R= 1$.
Then the geodesics satisfy $\ddot X_A =  X_A$(this equation of motion arises from the action for a point particle in embedding space subject to the constraint $X_A X^A = 1$) which means that
\be
X_A(s) = v_A \cosh (s)  + u_A \sinh(s)
\ee
for vectors $v_A$ and $v_A$ with $v_A u^A = 0$ and $v_A v^A - u_A u^A = 1$.  Note that
\be
y & =& \frac{1}{X_0 - X_3} \nn \\
z &=& \frac{X_z}{X_0 - X_3} \nn \\
\bar z &=& \frac{X_{\bar z}}{X_0 - X_3} 
\ee
so we end up with a simple formula for these coordinates on any geodesic.  Note that we have translation symmetry in $z, \bar z$ so we may as well set these to zero at a convenient point.  One choice is $z = \bar z = 0$ at $s=0$.  This means that $v_A$ will have vanishing $z, \bar z$ components.  A convenient Euclidean parameterization is
\be
X_0 &=& A_0 \cosh(s) + B_0 \sinh(s)
\nn \\
X_3 &=& A_3 \sinh(s) + B_3 \sinh(s)
\nn \\
X_z &=& A_z \sinh(s)
\nn \\
X_{\bar z} &=& A_{\bar z} \sinh(s)
\ee
We must have $B_3 = \frac{A_0 B_0}{A_3}$ and several other conditions for $B_0$ and $A_3$. Then if we set $A_0 = \frac{y_0}{2} + \frac{1}{2 y_0}$ then the point $s= z=\bar z = 0$ occurs at $y_0$.  Thus we find
\be
y(s) &=& y_0 \frac{e^s \left(y_0^2+z_0 \bar z_0\right)}{e^{2 s} z_0  \bar z_0+y_0^2}
\nn \\
z(s) &=& z_0 \frac{\left(1 - e^{2 s}\right) y_0^2 }{y_0^2 + e^{2 s} z_0 \bar z_0}
\nn \\
\bar z(s) &=& \bar z_0 \frac{\left(1 - e^{2 s}\right) y_0^2 }{y_0^2 + e^{2 s} z_0 \bar z_0}
\ee
Note that at $s = 0$ we have $(y_0,0,0)$ while for $s = -\infty$ we have $(0,z_0, \bar z_0)$.   We can also solve for $s$ in terms of $y$ or $z$, and then re-parameterize. It's simplest to solve for $s(z)$, which leads to
\be
\label{eq:GeodesicsinZ}
y(z) &=&  \sqrt{1 - \frac{z}{z_0}} \sqrt{y_0^2 + z \bar z_0}
\nn \\
\bar z(z) &=& \frac{\bar z_0}{ z_0} z
\ee
for geodesics beginning on the boundary at $z_0$ and ending in the bulk at $y_0$ and $z, \bar z = 0$.

\subsection{Global Reconstruction as a Boundary Operator Expansion}
\label{app:GlobalBOEBasics}

The ideas reviewed in this appendix were briefly explained in \cite{Paulos:2016fap}.   As far as we are aware, the explicit equations in this section were either first obtained by Miguel Paulos, or were derived by us via discussion and  collaboration with him.  Thus these results should largely be credited to Paulos and the other authors of \cite{Paulos:2016fap}.  A somewhat similar approach was taken in \cite{Nakayama:2015mva}.  Ultimately, the point is that the global conformal generators must act on $\phi$ as AdS isometries, and this  idea dates back to the beginning of AdS/CFT.  Throughout this appendix we will always be discussing the global $\phi$, which we will usually denote as $\phi^g$.

\subsubsection{Global BOE from HKLL Smearing}
\label{app:GlobalBOEfromHKLL}

Here we will show how to recover the global boundary operator expansion (BOE) for a scalar operator \cite{Goto:2017olq}	
\be
\label{eq:GlobalBOE}
\phi^{\text{g}}(y, z, \bar z) = y^{2h } \sum_{n=0}^\infty \frac{(-1)^n y^{2n}}{n! (2h)_n} \left( L_{-1} \bar L_{-1} \right)^n \CO(z, \bar z)
\ee
from the well-known HKLL \cite{Hamilton:2006az} smearing procedure.   

To obtain a free bulk scalar field from a boundary primary, we `smear' the boundary operator via
\be
\phi^{\text{g}}(y,0,0) = \frac{2h-1}{\pi} \int dz d \bar z \left( \frac{y^2 - z \bar z}{y} \right)^{2h-2} \CO(iz, i\bar z)
\ee
over the  Euclidean region $\bar z = z^*$ with $|z| < y$.  We can formally re-write this as
\be
\phi^{\text{g}}(y,0,0) = \frac{2h-1}{\pi} \int_{z \bar z<y^2} dz d \bar z \left( \frac{y^2 - z \bar z}{y} \right)^{2h-2} e^{iz \partial + i\bar z \bar \partial} \CO(0)
\ee
As the smearing function depends only on $z \bar z$, and terms with unequal powers of $z$ and $\bar z$ vanish after angular integration, we can change variables to
\be
\phi^{\text{g}}(y,0,0) = (2h-1)\int_0^{y^2} dx  \left( \frac{y^2 - x}{y} \right)^{2h-2}  P\left(- {x \partial \bar \partial} \right) \CO(0)  
\ee
where $P(a) = \sum_{n=0}^\infty \frac{1}{(n!)^2} a^n$.   One can do this integral explicitly and find a result with the desired series expansion in $y^2 \partial \bar \partial$.  One way to see this directly is  to perform a rescaling $x \to x y^2$ so that
\be
\phi^{\text{g}}(y,0,0) &=& (2h-1) y^{2h} \int_0^{1} dx \left( 1 - x \right)^{2h-2} P \left( -x \left( y^2 \partial \bar \partial \right) \right) \CO(0)  
\nn \\
&=& y^{2h} \sum_{n=0}^\infty \frac{(-1)^n}{(2h)_n n!} \left( y^2 \partial \bar \partial \right)^n \CO(0)
\ee
which is the desired  boundary operator expansion in powers of $y$.

\subsubsection{Bulk-Boundary Correlator from BOE}

Now we can verify explicitly that we obtain the correct $\< \phi \CO \>$ correlator from the boundary operator expansion for $\phi$.  In fact we will demonstrate a more general result, which makes it possible to compute $\< \phi^{\text{g}} \CO T(z_1) \cdots T(z_n) \>$:
\be
\< \phi^{\text{g}}(y,0,0) \CO(z, \bar z) T(z_1) \cdots T(z_n) \>  &=&  y^{2h} \sum_{n=0}^\infty \frac{(-1)^n (y^{2} \partial_x \partial_{\bar x})^n}{n! (2h)_n} \frac{f(z_i, x, z)}{(\bar z - \bar x)^{2h} }
\nn \\ &=&
\frac{y^{2h}}{\bar z^{2h}}   \sum_{n=0}^\infty \frac{1 }{n!} \left( -\frac{y^{2}}{\bar z} \partial_x \right)^n f(z_i, x, z)
\nn \\ &=&
\frac{y^{2h}}{\bar z^{2h}} f\left(z_i, -\frac{y^2}{\bar z}, z \right) 
\ee
where we define $f$ via $\< \CO(x) \CO(z) T(z_1) \cdots T(z_n) \> = f(z_i, x, z) (\bar z - \bar x)^{-2h}$, so we have
\be
\< \phi^{\text{g}} (y) \CO(z) T(z_1) \cdots T(z_n) \>   =  y^{2h}  \left\< \CO \left( -\frac{y^2}{\bar z} \right) \CO(z) T(z_1) \cdots T(z_n) \right\>
\ee
The simple special case of interest to us is
\be
\< \phi^{\text{g}}(y,0,0) \CO(z, \bar z) \> = \left( \frac{y}{y^2 + z \bar z} \right)^{2h}
\ee
as expected.

\subsubsection{Symmetries of the Global Boundary Operator Expansion}

In this section we will show that global conformal symmetry transformations $L_{-1}, L_0, L_1$ act as expected on the global conformally reconstructed $\phi$.

When we regard  $\phi$ as a bulk field, the global conformal generators should act on it as the differential operators 
\begin{align}
L_{-1} &= \partial_{z}
\nn \\
L_0 &= z \partial_{z} + \frac{1}{2} y \partial_y  
\nn \\
L_{1} &=  z^2 \partial_{z} + z y \partial_y -  y^2 \partial_{\bar z} 
\end{align}
So the goal is to show that when the quantum operators $L_n$ act on equation (\ref{eq:GlobalBOE}) in accord with this expectation.  In what follows, we will show that an $L_n$ transformation applied to $\CO$ results in the appropriate differential operator acting on $\phi$.

The fact that the translation generators act correctly follows easily because $\partial_z$ commutes with $(y^2 \partial \bar \partial)^n$.  For the dilatation $L_0$ note that
\be
\delta \phi^{\text{g}} &=& y^{2h} \sum_{n=0}^\infty \lambda_n y^{2n} (\partial \bar \partial )^n \left( z \partial + h \right) \CO(z, \bar z) 
\nn \\
&=& \left( z \partial + h \right) \phi^{\text{g}} + y^{2h} \sum_{n=0}^\infty n \, \lambda_n y^{2n} (\partial \bar \partial )^n  \CO(z, \bar z) 
\nn \\
&=& \left( z \partial + \frac{1}{2} y \partial_y  \right) \phi^{\text{g}}
\ee
as desired.   Note that this is automatic given the structure of expansion, and it does not depend on the form $\lambda_n = \frac{(-1)^n}{n! (2h)_n} $.

Finally, let us check the special conformal transformation $L_{1}$; we will see that it can only act appropriately if $\lambda_n$ take the expected form.  We need to compute
\begin{align*}
\label{eq:SCTonPhi}
\delta \phi^{\text{g}} =& y^{2h} \sum_{n=0}^\infty \lambda_n y^{2n} (\partial \bar \partial )^n \left( z^2 \partial +2 h z \right) \CO(z, \bar z) 
 \\
=& \left( z^2 \partial + 2h z \right)  \phi^{\text{g}} + y^{2h} \sum_{n=0}^\infty \lambda_n y^{2n} \left[ (\partial \bar \partial )^n,  z^2 \partial + 2h z \right] \CO(z, \bar z) 
\nn \\
= & z^{2}\partial\phi^{\text{g}}+zy\partial_{y}\phi^{\text{g}}+y^{2h}y^{2}\sum_{n=1}^{\infty}\lambda_{n}y^{2\left(n-1\right)}n\left(2h+n-1\right)\bar{\partial}^{n}\partial^{n-1}\mathcal{O}\left(z,\bar{z}\right)\\
= & z^{2}\partial\phi^{\text{g}}+zy\partial_{y}\phi^{\text{g}}+y^{2}\bar{\partial}y^{2h}\sum_{n=0}^{\infty}\lambda_{n+1}\left(n+1\right)\left(2h+n\right)y^{2n}\bar{\partial}^{n}\partial^{n}\mathcal{O}\left(z,\bar{z}\right)\\
= & \left(z^{2}\partial+zy\partial_{y}-y^{2}\bar{\partial}\right)\phi^{\text{g}}
\end{align*}
where in the last line, we used
\begin{equation}
\lambda_{n}=-\lambda_{n+1} (n+1)(2h+n).
\end{equation}
The same result could also be obtained by demanding that the conformal Casimir acts appropriately on $\phi^{\text{g}}$, as shown by M. Paulos.

\section{Regulation: from Classical Backgrounds to  Correlators}
\label{subsec:FromClassicalBackgroundtoMultiTVacuumCorrelators}

In section \ref{sec:GravitationalWilsonLines}, we developed an algorithm to compute the correlators  $\langle T\dots T\bar{T}\dots\bar{T}\phi \CO\rangle$ from
the simpler correlator $\langle\phi \CO\rangle_{\mu,\bar{\mu}}$ evaluated in states with non-trivial stress tensor vevs:
\begin{equation}
\langle T(z)\rangle_{\mu,\bar{\mu}}=T_{cl}\left(\bar{z}\right),\hspace{1em}\langle\bar{T}(z)\rangle_{\mu,\bar{\mu}}=\bar{T}_{cl}\left(\bar{z}\right)
\end{equation}
The algorithm was to first view $\langle\phi \CO\rangle_{\mu,\bar{\mu}}$ as a functional on the vevs $T_{cl}(z)$ and $\bar{T}_{cl}(\bar{z})$. In a series
expansion, this functional takes the general form:
\begin{footnotesize}
\begin{align}
\langle\phi \CO\rangle_{\mu,\bar{\mu}}=&\langle\phi \CO\rangle_{0}\left(1+\int dx\tilde{K}_{10}(x)T_{cl}(x)+\int d\bar{x}\tilde{K}_{01}(\bar{x})\bar{T}_{cl}(\bar{x})+\int dx\tilde{K}_{11}(x,\bar{x})T_{cl}(x)\bar{T}_{cl}(\bar{x})+\dots\right)\nn\\
=&\langle\phi \CO\rangle_{0}\left(\sum_{n,\bar{n}=0}^{\infty}\int\prod_{i=1}^{n}dx_{i}\prod_{\bar{i}=1}^{\bar{n}}d\bar{x}_{\bar{i}}\tilde{K}_{i,\bar{i}}(x_{1},\dots x_{n},\bar{x}_{1},\dots,\bar{x}_{n})T_{cl}(x_{1})\dots T_{cl}(x_{n})\bar{T}_{cl}(\bar{x}_{1})\dots\bar{T}_{cl}(\bar{x}_{\bar{n}})\right)\label{eq:ExpandphiO2}
\end{align}
\end{footnotesize}
Then we compute the vacuum sector of the operator product $\phi \CO$ that includes all contributions from Virasoro descendants of the vacuum\footnote{All other contributions to $\phi O$ involve quantum operators that are not descendants of the vacuum. Thus they do not contribute to the multi-T correlators that we are computing in this appendix.}, which is done by replacing $T_{cl}$ and $\bar{T}_{cl}$ in $\<O\phi\>_{\mu,\bar{\mu}}$ by quantum operators $T$ and $\bar{T}$. 

However, generically operators products of  $T$ have short distance singularities when two $T$'s approach each other, which will occur due to the integration over positions in (\ref{eq:ExpandphiO2}). In \cite{Fitzpatrick:2016mtp} we empirically discovered a simple regulator (equation C.10 there) that, when applied to the ``quantum" version of (\ref{eq:ExpandphiO2}), produces the correct OPE block. 
The correlator between the regulated product of $T$'s, denoted as $[T(x_{1})\dots T(x_{n})]$, and external, unregulated $T(z_{i})$s were found to be:
\begin{equation}
\langle T(z_{1})\dots T(z_{k})[T(x_{1})\dots T(x_{n})]\rangle=0,\hspace{1em}n>k\label{eq:C10Regulator1}
\end{equation}
\begin{equation}
\langle T(z_{1})\dots T(z_{k})[T(x_{1})\dots T(x_{n})]\rangle\equiv\sum_{groupings}\prod_{i=1}^{n}\langle T(z_{1})\dots T(z_{k})T(x_{i})\rangle,\hspace{1em}n\le k\label{eq:C10Regulator}
\end{equation}
The sum is over different groupings of $T(z_{i})$'s. Note that since in each correlator there is only one $T(x_{i})$, the results never diverge as $x_{i}\rightarrow x_{j}$.  Thus the regulator fully specifies correlators of the OPE block with stress tensors. 

To summarize, we proposed that the vacuum sector of the $\phi \CO$ operator product is:
\begin{equation}
\phi \CO=\left[\left.\langle\phi \CO\rangle_{B}\right|_{T_{cl}\rightarrow T,\bar{T}_{cl}\rightarrow\bar{T}}\right]+\dots
 \label{eq:OPEBlockSummary}
\end{equation}
where the square bracket represents the regularization applied to all products of $T$ and $\bar{T}$'s. In the current context this regulator is defined by (\ref{eq:C10Regulator1}-\ref{eq:C10Regulator}). In \cite{Fitzpatrick:2016mtp} and this paper, this proposal survived extensive and non-trivial checks by direct computation. 

In this appendix, we would like to provide a general argument for this proposal.  In particular, we would like to show that, under fairly
general assumptions, it correctly extracts multi-T vacuum correlators such as $\langle T(z_1)\dots T(z_n) \bar{T}(\bar{z}_1)\dots \bar{T}(\bar{z}_{\bar{n}})\phi \CO\rangle_{0}$ from simpler core correlators such as $\langle\phi O\rangle_{\mu,\bar{\mu}}$ on a background with non-trivial source. We also show that this algorithm does not seem to rely on conformal symmetry and may work in a wider range of settings. 

Suppose we have a generic field theory containing a bosonic quantum
operator $T$. It is possible to construct a classical source for
it, such that $T$ has a classical vev:
\begin{equation}
\langle T(x)\rangle_{\mu}=T_{cl}(x).\label{eq:muToT}
\end{equation}
We view this equation as a mapping between functions $\mu\leftrightarrow T_{cl}$.
We will make the assumption this mapping is one-to-one, and $\mu=0$ maps to $T_{cl}=0$. In
particular, this assumes that given any $T_{cl}(x)$, there must exist a unique source configuration $\mu(x)$ that sets up this vev. Thus we can write the functional $\mu[T_{cl}]$ as the solution of (\ref{eq:muToT}). Note that the source is defined in the usual way by shifting the action in the Euclidean path integral:
\begin{equation}
S\rightarrow S+\int dz\mu(z)T(z)
\end{equation}
The input of our algorithm is $\langle X\rangle_{\mu[T_{cl}]}$ as a functional on $T_{cl}$. 
\begin{equation}
\langle X\rangle_{\mu[T_{cl}]}=\langle Xe^{\int dz\mu[T_{cl}](z)T(z)}\rangle_{0}
\end{equation}
Once this is known, we should have enough information to determine
vacuum multi-T correlators $\langle XT(z_{1})\dots T(z_{n})\rangle_{0}$.
We first compute the simplest of this family:
\be
\langle XT(z_{1})\rangle_{c,0} &=&\left.\frac{\delta}{\mu(z_{1})}\langle X\rangle_{\mu}\right|_{\mu\rightarrow0}  \nn \\
 & =&\int dx_{1}\left.\frac{\delta T_{cl}(x_{1})}{\mu(z_{1})}\frac{\delta}{\delta T_{cl}(x_{1})}\langle X\rangle_{\mu[T_{cl}]}\right|_{\mu\rightarrow0} \nn \\
 & =&\langle X\rangle_{0}\int dx_{1}\langle T(z_{1})T(x_{1})\rangle_{0}\tilde{K}_{10}^{X}(x_{1}) \nn \\
 & =&\langle T(z_{1})\left[\langle X\rangle_{\mu[T]}\right]\rangle_{0}\label{eq:1TFromBackground}
\ee
In the second step, we used: 
\be
\left.\frac{\delta T_{cl}(x_{1})}{\mu(z_{1})}\right|_{\mu\rightarrow0} & =&\left.\frac{\delta}{\mu(z_{1})}\langle T(x_{1})\rangle_{\mu}\right|_{\mu\rightarrow0}=\langle T(z_{1})T(x_{1})\rangle_{0}\\
\left.\frac{\delta\langle X\rangle_{\mu_{[T_{cl}]}}}{\delta T_{cl}(x_{1})}\right|_{\mu\rightarrow0} & =&\left.\frac{\delta}{\delta T_{cl}(x_{1})}\langle X\rangle_{0}\int dx\tilde{K}_{1}^{X}(x)T_{cl}(x)\right|_{\mu\rightarrow0}=\langle X\rangle_{0}\tilde{K}_{1}^{X}(x_{1})
\ee
where we have inserted a series expansion of $\langle X\rangle_{\mu_{[T_{cl}]}}$
in the style of (\ref{eq:ExpandphiO2}), which should exist given the
non-singular limit $\langle X\rangle_{\mu[T_{cl}\rightarrow0]}=\langle X\rangle_{0}$.
When we replace $X\rightarrow \phi O$, (\ref{eq:1TFromBackground}) is precisely the result predicted by inserting (\ref{eq:OPEBlockSummary}) into $\langle T(z_1) \phi \CO\rangle$ and evaluate using (\ref{eq:C10Regulator1}-\ref{eq:C10Regulator}). We made this clear in the last step.

A slightly more non-trivial example is $\langle XTT\rangle$: 
\be
\langle XT(z_{1})T(z_{2})\rangle_{c,0} &=&\left.\frac{\delta}{\delta\mu(z_{1})}\frac{\delta}{\delta\mu(z_{2})}\langle X\rangle_{\mu}\right|_{\mu\rightarrow0} \nn \\
 & =&\int dx_{2}\left.\frac{\delta}{\delta\mu(z_{1})}\left(\frac{\delta T_{cl}(x_{2})}{\delta\mu(z_{2})}\frac{\delta}{\delta T_{cl}(x_{2})}\langle X\rangle_{\mu[T_{cl}]}\right)\right|_{T_{cl}\rightarrow0} \nn \\
 & =&\left.\int dx_{2}\frac{\delta^{2}T_{cl}(x_{2})}{\delta\mu(z_{1})\delta\mu(z_{2})}\frac{\delta}{\delta T_{cl}(x_{2})}\langle X\rangle_{\mu[T_{cl}]}\right|_{T_{cl}\rightarrow0} \nn \\
 && +\left.\int dx_{1}dx_{2}\frac{\delta T_{cl}(x_{2})}{\delta\mu(z_{2})}\frac{\delta T_{cl}(x_{1})}{\delta\mu(z_{1})}\frac{\delta}{\delta T_{cl}(x_{1})}\frac{\delta}{\delta T_{cl}(x_{2})}\langle X\rangle_{T_{cl}}\right|_{T_{cl}\rightarrow0} \nn  \\
 & =&\langle X\rangle_{0}\int dx_{1}\langle T(z_{1})T(z_{2})T(x_{1})\rangle\tilde{K}_{1}^{X}(x_{1}) \nn \\
 && +\langle X\rangle_{0}\int dx_{1}dx_{2}\langle T(z_{1})T(x_{1})\rangle_{0}\langle T(z_{2})T(x_{2})\rangle_{0}\tilde{K}_{2}^{X}(x_{1},x_{2}) \nn \\
 & =&\langle T(z_{1})T(z_{2})\left[\langle X\rangle_{\mu[T]}\right]\rangle_{0}
\ee
Again this exactly agrees with the result of our OPE block defined with regulator (\ref{eq:C10Regulator}). It is easy to see why this works to level $n$, $\langle XT(z_{1})\dots T_{n}(z_{n})\rangle$:
\begin{equation}
\langle XT(z_{1})\dots T(z_{n})\rangle_{0}=\left.\frac{\delta}{\mu(z_{1})}\dots\frac{\delta}{\mu(z_{n})}\langle X\rangle_{\mu}\right|_{\mu\rightarrow0}
\end{equation}
Each time we add a $T(z_{n+1})$, the corresponding $\frac{\delta}{\delta\mu(z_{n+1})}$
either act on $\langle \CO\CO\rangle_{T_{cl}}$ as $\int dx_{n+1}\frac{\delta T(x_{n+1})}{\delta\mu(z_{n+1})}\frac{\delta}{\delta T(x_{n+1})}$,
where it picks up a single $T$ from the OPE block of $\CO\CO$, or acts
on an existing derivative $\frac{\delta^{k}T_{cl}(x_{i_{k}})}{\delta\mu(z_{i_{1}})\dots\delta\mu(z_{i_{k}})}\frac{\delta}{\delta T_{cl}(x_{i_{k}})}$, where it adds a point to a existing multi-T correlator. By construction, there are never two $T$'s from the $X$ OPE block appearing in the same vev. Thus, there are no UV divergences. The result is  our OPE block defined with regulator (\ref{eq:C10Regulator}). 

To summarize, given the correlator of operator product $X$ on non-trivial backgrounds, $\langle X\rangle_{\mu[T_{cl}]}$, we can extract the vacuum correlator between $X$
and any number of $T$ insertions using: 
\begin{align}
\langle T(z_{1})\dots T(z_{n})X\rangle_{0} & =\sum_{groupings}\left.\prod_{i<n}\int dx_{n}\langle T(z_{i_{1}})\dots T(z_{i_{k_{n}}})T(x_{i})\rangle\frac{\delta}{\delta T_{cl}(x_{i})}\langle X\rangle_{\mu[T_{cl}]}\right|_{T_{cl}\rightarrow0} \nn \\
 & =\langle T(z_{1})\dots T(z_{n})\left[\langle X\rangle_{\mu[T]}\right]\rangle_{0}
\end{align}
where in the second line we interpreted the result as computing the correlator between $T(z_{1})\dots T(z_{n})$ and the OPE block of
the operator product $X$, which is constructed and regulated as given in the first line. This algorithm should work in any field theory
as long as the mapping $\langle T\rangle_{\mu}=T_{cl}$ is one-to-one between $\mu$ and $T_{cl}$.

\section{Bulk Virasoro Transformations}
\label{app:BulkVirasoroTransforms}

We would like to find an extension of a boundary Virasoro transformation into the bulk, such that this bulk transformation will preserve the Fefferman-Graham form of the metric. To achieve this, this bulk Virasoro transformation must depend on the initial bulk metric. In other words, the Virasoro transformations acts in the following way:
\begin{equation}
\left(z,\bar{z},y,f,\bar{f}\right)\rightarrow\left(\tilde{z},\tilde{\bar{z}},\tilde{y},\tilde{f},\tilde{\bar{f}}\right)
\end{equation}
The bulk metric is specified by $\left(f\left(z\right),\bar{f}\left(\bar{z}\right)\right)$,
which determines the vev of stress tensors and the boundary Virasoro transformations back to the uniformizing coordinate (\ref{eq:GeneralVacDiffeomorphism}), reproduced here:
\begin{equation}
z_{u}=f\left(z\right)-\frac{2y^{2}f^{\prime2}\bar{f}^{\prime\prime}}{4f^{\prime}\bar{f}^{\prime}+y^{2}f^{\prime\prime}\bar{f}^{\prime\prime}},\hspace{1em}\bar{z}_{u}=\bar{f}\left(\bar{z}\right)-\frac{2y^{2}\bar{f}^{\prime2}f^{\prime\prime}}{4f^{\prime}\bar{f}^{\prime}+y^{2}f^{\prime\prime}\bar{f}^{\prime\prime}}
\end{equation}
\begin{equation}
y_{u}=4y\frac{\left(f^{\prime}\bar{f}^{\prime}\right)^{\frac{3}{2}}}{4f^{\prime}\bar{f}^{\prime}+y^{2}f^{\prime\prime}\bar{f}^{\prime\prime}}
\end{equation}

Collectively, we may denote $P=\left(z,\bar{z},y,f,\bar{f}\right)$ and the above coordinate map to the uniformizing coordinate is denoted as $P_{u}\left(P\right)$.
Given any Virasoro transformation $\left(g(z),\bar{g}\left(\bar{z}\right)\right)$, the way we obtain its bulk completion on any background metric that preserves the Fefferman-Grahm gauge is to first map the original coordinate back to the uniformizing coordinate, and then transform from it to the new coordiante such that the composition is equivalent to $\left(g(z),\bar{g}\left(\bar{z}\right)\right)$ on the boundary. In equations, this means the new point in the $\tilde{P}$ satisfies
\begin{equation}
P_{u}\left(\tilde{P}\right)=P_{u}\left(P\right)
\end{equation}
\begin{equation}
\tilde{f}^{-1}\circ f\left(z\right)=g\left(z\right),\hspace{1em}\tilde{\bar{f}}^{-1}\circ\bar{f}\left(\bar{z}\right)=\bar{g}\left(\bar{z}\right)
\end{equation}

We consider a generic background that is specified by $\left(f\left(z\right),\bar{f}\left(\bar{z}\right)\right)$. Then we do a small Virasoro transformation generated by $L_m$ on this background. On the boundary, this transformation is defined as
\begin{equation}
\left(1+\epsilon L_{m}\right)z=z+\epsilon z^{m+1}
\end{equation}
This transformation takes 
\begin{equation}
\left(f,\bar{f}\right)\rightarrow\left(\tilde{f},\bar{f}\right)
\end{equation}
$\tilde{f}$ is determined by:
\begin{equation}
f^{-1}\circ\tilde{f}\left(z\right)=z-\epsilon z^{m+1}
\end{equation}
which means
\begin{equation}
\tilde{f}=f-\epsilon z^{m+1}f^{\prime}\equiv f+\epsilon\delta_m f
\end{equation}
We then solve
\begin{equation}
P_{u}\left(P+\epsilon\delta_m P\right)=P_{u}\left(P\right)
\end{equation}
The solution is (\ref{eq:LmTransformations}), reproduced here:
\be
\delta_{m}z&=&\frac{z^{m-1}\left(\left(m^{2}+m+z^{2}S(z)\right)\bar{S}\left(\bar{z}\right)y^{4}-4z^{2}\right)}{y^{4}S(z)\bar{S}\left(\bar{z}\right)-4}\\
\delta_{m}\bar{z}&=&\frac{2m(m+1)y^{2}z^{m-1}}{y^{4}S(z)\bar{S}\left(\bar{z}\right)-4}\\
\delta_{m}y&=&\frac{1}{2}(m+1)yz^{m}
\ee
Note that the $f$ and $\bar{f}$ organize themselves exactly to reproduce
$S$ and $\bar{S}$, where 
\begin{equation}
\bar{S}=\frac{\bar{f}^{(3)}\bar{f}'-\frac{3}{2}\bar{f}''^{2}}{\bar{f}'^{2}}=\frac{12}{c}\bar{T}
\end{equation}
Clearly, we see that $L_{m}$  with $m\ge2$ will leave points $(y,0,0)$ invariant.

For $L_{1}$ this is explicitly not the case. In fact, the action
of $L_{1}$ is somewhat non-trivial. On a background with $L=0$ (correlators $\<\phi\CO \bar T\cdots \bar T\>$ without any $T$),
we have: 
\begin{equation}
L_{1}\phi\left(y,0,0\right)=\left(-y^{2}\bar{\partial}-\frac{6}{c}y^{4}\bar{T}\left(0\right)\partial\right)\phi\left(y,0,0\right).
\label{eq:L1phiAntiChiral}
\end{equation}
One way to test whether this is correct is to compute 
\begin{equation}
\langle \CO(z,\bar{z})\bar{T}(\bar{z}_{1})L_{1}\phi\left(y,0,0\right)\rangle\stackrel{?}{=}\langle \CO(z,\bar{z})\bar{T}(\bar{z}_{1})\left(-y^{2}\bar{\partial}-\frac{6}{c}y^{4}\bar{T}\left(0\right)\partial\right)\phi\left(y,0,0\right)\rangle
\end{equation}
Note that the first term on the RHS, which is the naive transformation of $\phi$ (it's the transformation of $\phi^\text{global}$ under $L_1$),
gives a wrong result: 
\begin{align}
&-y^{2}\langle \CO(z,\bar{z})\bar{T}(\bar{z}_{1})\bar{\partial}\phi\left(y,0,0\right)\rangle\\  
=&\frac{2hy^{2}\bar{z}\left(y^{2}z\bar{z}\left(2(h-3)\bar{z}_{1}\bar{z}-3(h-1)\bar{z}_{1}^{2}+3\bar{z}^{2}\right)-z^{2}\bar{z}^{2}\bar{z}_{1}\left((h-1)\bar{z}_{1}+\bar{z}\right)+3y^{4}\left(\bar{z}-\bar{z}_{1}\right){}^{2}\right)}{\left(\bar{z}-\bar{z}_{1}\right){}^{2}\bar{z}_{1}^{4}\left(z\bar{z}+y^{2}\right)^{2}}\nn
\end{align}
This is wrong because it has a $\frac{1}{\bar{z}_{1}^{4}}$ pole, which is inconsistent with the condition of equation (\ref{eq:Conditions}). But the second term 
\begin{equation}
-\frac{6}{c}y^{4}\langle \CO(z,\bar{z})[\bar{T}(\bar{z}_{1})\bar{T}\left(0\right)]\partial\phi\left(y,0,0\right)\rangle
=-3y^{4}\frac{1}{\bar{z}_{1}^{4}}\langle \CO(z,\bar{z})\partial\phi\left(y,0,0\right)\rangle=-\frac{6hy^{4}\bar{z}}{\bar{z}_{1}^{4}\left(z\bar{z}+y^{2}\right)}
\end{equation}
has precisely the right form to cancel this pole. Then combining these two terms, we have
\begin{align}
&\langle \CO(z,\bar{z})\bar{T}(\bar{z}_{1})\left(-\frac{6}{c}y^{4}\bar{T}\left(0\right)\partial-y^{2}\bar{\partial}\right)\phi\left(y,0,0\right)\rangle\nn\\
=&\langle \CO(z,\bar{z})\phi\left(y,0,0\right)\rangle\frac{2hy^{2}z\bar{z}^{2}\left(h\left(y^{2}\left(2\bar{z}-3\bar{z}_{1}\right)-z\bar{z}\bar{z}_{1}\right)+z\bar{z}\left(\bar{z}_{1}-\bar{z}\right)\right)}{\left(\bar{z}-\bar{z}_{1}\right){}^{2}\bar{z}_{1}^{3}\left(z\bar{z}+y^{2}\right){}^{2}}\nn\\
=&-(2h z + z^{2} \partial_{z})\langle O(z,\bar{z})\bar{T}(\bar{z}_{1})\phi\left(y,0,0\right)\rangle\\
=&-\langle\left[L_{1},\CO(z,\bar{z})\right]\bar{T}(\bar{z}_{1})\phi\left(y,0,0\right)\rangle\nn\\
=&\langle \CO(z,\bar{z})\bar{T}(\bar{z}_{1})L_{1}\phi\left(y,0,0\right)\rangle\nn
\end{align}
Similarly, we checked (\ref{eq:L1phiAntiChiral}) also work in the case of $\langle  \bar{T}\bar T \CO L_1 \phi\rangle$. In particular, we checked that 
\begin{align}
&\langle\bar{T}(\bar{z}_{1})\bar{T}(\bar{z}_{2})\CO(z,\bar{z})L_{1}\phi(y,0,0)\rangle\\
=&\langle\bar{T}(\bar{z}_{1})\bar{T}(\bar{z}_{2})\CO(z,\bar{z})\left(-y^{2}\bar{\partial}-\frac{6}{c}\bar{T}(0)y^{4}\partial\right)\phi(y,0,0)\rangle\\
=&-y^{2}\langle\bar{T}(\bar{z}_{1})\bar{T}(\bar{z}_{2})\CO(z,\bar{z})\bar{\partial}\phi(y,0,0)\rangle\\
&-\frac{6}{c}y^{4}\langle\bar{T}(\bar{z}_{1})\bar{T}(\bar{z}_{2})\bar{T}(0)\rangle\langle \CO(z,\bar{z})\partial\phi(y,0,0)\rangle\\
&-\frac{6}{c}y^{4}\left(\langle\bar{T}(\bar{z}_{1})\bar{T}(0)\rangle\langle\bar{T}(\bar{z}_{2})\CO(z,\bar{z})\partial\phi(y,0,0)\rangle+\left(z_1\leftrightarrow z_2\right)\right)\\
=&-\left(2hz+z^{2}\partial_{z}\right)\langle\bar{T}(\bar{z}_{1})\bar{T}(\bar{z}_{2})\CO(z,\bar{z})\phi(y,0,0)\rangle\\
=&-\langle\bar{T}(\bar{z}_{1})\bar{T}(\bar{z}_{2})\left[L_{1},\CO(z,\bar{z})\right]\phi(y,0,0)\rangle.
\end{align}
The fact that these work nicely are non-trivial checks for our method.

\section{Additional Technical Results}

\subsection{Gravitational Wilson Line Computations at Higher Orders}
\label{app:GravitationalComputationsHigherOrders}

In this section, we provide the details to derive the bulk-boundary
OPE block kernels up to order $\frac{1}{c^{2}}$.  First, we need to solve the following equation at large $c$ 
\be
S\left(f,z\right)\equiv\frac{f'''\left(z\right)f'\left(z\right)-\frac{3}{2}\left(f''\left(z\right)\right)^{2}}{\left(f'\left(z\right)\right)^{2}}=\frac{12}{c}T\left(z\right)
\ee
and determine $f\left(z\right)$ and $\overline{f}\left(\overline{z}\right)$
as functions of the stress tensor operators $T\left(z\right),\overline{T}\left(\overline{z}\right)$.
We'll do this by expanding $f\left(z\right)$ in terms of large $c$
as follows
\be\label{eq:fLargeCExpansion}
f\left(z\right)=f_{0}\left(z\right)+\sum_{n=1}^{\infty}\frac{f_{n}\left(z\right)}{c^{n}}
\ee
with $f_{0}\left(z\right)=z$ satisfies $S\left(f_{0}\left(z\right),z\right)=0$
at leading order. At order $\frac{1}{c}$ and $\frac{1}{c^{2}}$,
$f_{1}\left(z\right)$ and $f_{2}\left(z\right)$ are determined by the following differential equations
\begin{align}\label{eq:f1f2Equations}
f_{1}^{\left(3\right)}\left(z\right)-12T\left(z\right) & =0,\\
2f_{1}{}^{(3)}(z)f_{1}'(z)+3f_{1}''(z){}^{2}-2f_{2}{}^{(3)}(z) & =0.\nn
\end{align}
The first equation is easy to solve and the solution with desired
boundary condition is 
\be
f_{1}\left(z\right)=6\int_{0}^{z}dz'\left(z-z'\right)^{2}T\left(z'\right).
\ee
Using this solution, the second equation in (\ref{eq:f1f2Equations}) becomes 
\begin{align*}
f_{2}^{\left(3\right)} & =f_{1}^{\left(3\right)}f_{1}^{'}+\frac{3}{2}f_{1}^{''2}\\
 & =144T\left(z\right)\int_{0}^{z}dz'\left(z-z'\right)T\left(z'\right)+432\int_{0}^{z}dz'\int_{0}^{z'}dz'''T\left(z'\right)T\left(z'''\right).
\end{align*}
And the solution is 
\begin{align*}
f_{2}\left(z\right) & =36\int_{0}^{z}dz''\left(z-z''\right)^{2}\left[2T\left(z''\right)\int_{0}^{z''}dz'\left(z''-z'\right)T\left(z'\right)+6\int_{0}^{z''}dz'\int_{0}^{z'}dz'''T\left(z'\right)T\left(z'''\right)\right]\\
 & =72\int_{0}^{z}dz'\int_{0}^{z'}dz''T\left(z'\right)T\left(z''\right)\left(z-z'\right)^{2}\left(z-z''\right).
\end{align*}

Now we can turn to the derivation of the bulk-boundary OPE block kernels.
Expanding the coordinates transformation (\ref{eq:GeneralVacDiffeomorphism}) in terms of large $c$, i.e.
using \ref{eq:fLargeCExpansion} with $f_{0}\left(z\right)=z$, we have  
\begin{small}
\begin{align*}
u & =y+\frac{y\left(\bar{f}_{1}'\left(\bar{z}\right)+f_{1}'(z)\right)}{2c}-\frac{y\left(2y^{2}f_{1}''(z)\bar{f}_{1}''\left(\bar{z}\right)+\left(f_{1}'(z)-\bar{f}_{1}'\left(\bar{z}\right)\right){}^{2}-4\left(\bar{f}_{2}'\left(\bar{z}\right)+f_{2}'(z)\right)\right)}{8c^{2}}+\mathcal{O}\left(c^{-3}\right)\\
w & =z+\frac{f_{1}(z)-\frac{1}{2}y^{2}\bar{f}_{1}''\left(\bar{z}\right)}{c}+\frac{2f_{2}(z)-y^{2}\left(\left(f_{1}'(z)-\bar{f}_{1}'\left(\bar{z}\right)\right)\bar{f}_{1}''\left(\bar{z}\right)+\bar{f}_{2}''\left(\bar{z}\right)\right)}{2c^{2}}+\mathcal{O}\left(c^{-3}\right)
\end{align*}
\end{small}
and similar expression for $\overline{w}$. Expanding the bulk-boundary
two-point function and using the above result, we find
\begin{align*}
 & \log\phi\left(y,z_{f},z_{f}\right)\mathcal{O}\left(z_{i},\overline{z}_{i}\right)\\
= & 2h\log\left(\frac{u_{f}\sqrt{w'\left(z_{i}\right)\overline{w}\left(\overline{z}_{i}\right)}}{u_{f}^{2}+\left(w_{f}-w_{i}\right)\left(\overline{w}_{f}-\overline{w}_{i}\right)}\right)\\
= & 2h\log\left(\frac{y}{y^{2}+z\bar{z}}\right)+\underbrace{\frac{h}{c}\frac{\left(z\bar{z}+y^{2}\right)f_{1}'(z)-2\bar{z}f_{1}(z)}{\left(z\bar{z}+y^{2}\right)}}_{K_{T}}\underbrace{-\frac{2hy^{2}f_{1}(z)\bar{f}_{1}\left(\bar{z}\right)}{c^{2}\left(z\bar{z}+y^{2}\right)^{2}}}_{K_{T\overline{T}}}\\
 & \underbrace{-\frac{h\left(\left(z\bar{z}+y^{2}\right)\left(\left(z\bar{z}+y^{2}\right)\left(f_{1}'(z){}^{2}-2f_{2}'(z)\right)+4\bar{z}f_{2}(z)\right)-2\bar{z}^{2}f_{1}(z){}^{2}\right)}{2c^{2}\left(z\bar{z}+y^{2}\right){}^{2}}}_{K_{TT}}\\
 & +K_{\overline{T}}+K_{\overline{T}\overline{T}}+\mathcal{O}\left(c^{-3}\right)
\end{align*}
with $K_{\overline{T}}$ and $K_{\overline{T}\overline{T}}$ the complex
conjugate of $K_{T}$ and $K_{TT}$ respectively. In the third line
of the above equations, we've put the two operators at $\phi\left(y,0,0\right)$
and $\mathcal{O}\left(z,\overline{z}\right)$. 

Plugging in the solutions for $f_{n},\overline{f}_{n}$, we have
\begin{align*}
K_{T} & =\frac{12h}{c}\int_{0}^{z}dz'\frac{\left(y^{2}+z'\overline{z}\right)\left(z-z'\right)}{z\bar{z}+y^{2}}T\left(z'\right)\\
K_{TT} & =\int_{0}^{z}dz'\int_{0}^{z'}dz''\frac{72h\left(z-z'\right)^{2}\left(y^{2}+\overline{z}z''\right)^{2}}{c^{2}\left(z\bar{z}+y^{2}\right){}^{2}}T\left(z'\right)T\left(z''\right)\\
K_{T\overline{T}} & =-\frac{72hy^{2}}{c^{2}\left(z\bar{z}+y^{2}\right){}^{2}}\int_{0}^{z}dz'\left(z-z'\right)^{2}\int_{0}^{\overline{z}}d\overline{z}'\left(\overline{z}-\overline{z}'\right)^{2}T\left(z'\right)\overline{T}\left(\overline{z}'\right)
\end{align*}
Sending $y=0$, we find 
\begin{align*}
K_{T} & \xRightarrow{y=0}\frac{12h}{c}\int_{0}^{z}dz'\frac{z'\left(z-z'\right)}{z}T\left(z'\right)\\
K_{TT} & \xRightarrow{y=0}\int_{0}^{z}dz'\int_{0}^{z'}dz''\frac{72h\left(z-z'\right)^{2}z''^{2}}{c^{2}z{}^{2}}T\left(z'\right)T\left(z''\right)\\
K_{T\overline{T}} & \xRightarrow{y=0}0
\end{align*}
which are exactly the boundary-boundary OPE kernels found in \cite{Fitzpatrick:2016mtp}.

\subsection{Computations Using the Bulk-Boundary OPE Block}
\label{app:ComputationsBBOPEBlock}
In this section, we'll provide the details for computing $\left\langle \phi\mathcal{O}TT\right\rangle $
and $\left\langle \phi\mathcal{O}T\overline{T}\right\rangle $ using
bulk-boundary OPE block with the regulator proposed in Appendix C.2 of \cite{Fitzpatrick:2016mtp} and discussed in details in appendix \ref{subsec:FromClassicalBackgroundtoMultiTVacuumCorrelators}. The regulator (\ref{eq:C10Regulator1}-\ref{eq:C10Regulator}) is basically saying that when computing $\<\phi\CO T_1\cdots T_n\bar T_1\cdots\bar T_m \>$, the kernels in the OPE block of $\phi\CO$ that will contribute are those whose numbers of $T$ and $\bar T$ are equal or less than $n$ and $m$ respectively. 

\subsubsection{$\left\langle \phi\mathcal{O}T\overline{T}\right\rangle $ }

Using the regulator (\ref{eq:C10Regulator1}-\ref{eq:C10Regulator}), the kernels in the bulk-boundary OPE of $\phi\CO$ that contribute to $\left\langle \phi\mathcal{O}T\overline{T}\right\rangle $ are $K_TK_{\bar T}$ and $K_{T\bar T}$. So $\left\langle \phi\mathcal{O}T\overline{T}\right\rangle $ 
is given by
\begin{align}
\frac{\left\langle \phi\left(y,0,0\right)\mathcal{O}\left(z,\overline{z}\right)T\left(z_{1}\right)\overline{T}\left(\overline{w}_{1}\right)\right\rangle }{\left\langle \phi\left(y,0,0\right)\mathcal{O}\left(z,\overline{z}\right)\right\rangle }= & \left\langle e^{K_{T}+K_{\overline{T}}+K_{T\overline{T}}+\cdots}T\left(z_{1}\right)\overline{T}\left(\overline{w}_{1}\right)\right\rangle \nonumber \\
= & \left\langle \left(K_{T}K_{\overline{T}}+K_{T\overline{T}}\right)T\left(z_{1}\right)\overline{T}\left(\overline{w}_{1}\right)\right\rangle \label{eq:PhiOTTbarUsingOPEBlock}
\end{align}
The first term is
\begin{align*}
 & \left\langle K_{T}K_{\overline{T}}T\left(z_{1}\right)\overline{T}\left(\overline{w}_{1}\right)\right\rangle \\
= & \left(\frac{144h^{2}}{c^{2}}\int_{0}^{z}dz'\int_{0}^{\overline{z}}d\overline{z}'\frac{\left(y^{2}+z'\overline{z}\right)\left(z-z'\right)}{z\bar{z}+y^{2}}\frac{\left(y^{2}+\overline{z}'z\right)\left(\overline{z}-\overline{z}'\right)}{z\bar{z}+y^{2}}\right)\left\langle \left[T\left(z'\right)\overline{T}\left(\overline{z}'\right)\right]T\left(z_{1}\right)\overline{T}\left(\overline{w}_{1}\right)\right\rangle \\
= & \frac{h^{2}z^{2}\bar{z}^{2}\left(y^{2}\left(3\bar{w}_{1}-2\bar{z}\right)+\bar{w}_{1}z\bar{z}\right)\left(y^{2}\left(3z_{1}-2z\right)+z_{1}z\bar{z}\right)}{z_{1}^{3}\overline{w}_{1}^{3}\left(z-z_{1}\right){}^{2}\left(\bar{w}_{1}-\bar{z}\right){}^{2}\left(z\bar{z}+y^{2}\right){}^{2}}
\end{align*}
where in the second line we use the regulated four-point function 
\be
\left\langle \left[T\left(z'\right)\overline{T}\left(\overline{z}'\right)\right]T\left(z_{1}\right)\overline{T}\left(\overline{w}_{1}\right)\right\rangle =\frac{c^{2}}{4}\frac{1}{\left(z'-z_{1}\right)^{4}\left(\overline{z}'-\overline{w}_{1}\right)^{4}}.
\ee
The above result is just the contribution from $\left\langle \phi\mathcal{O}T\right\rangle \left\langle \phi\mathcal{O}\overline{T}\right\rangle $.

The second term in equation (\ref{eq:PhiOTTbarUsingOPEBlock}) is
\begin{align}
 & \left\langle K_{T\overline{T}}T\left(z_{1}\right)\overline{T}\left(\overline{w}_{1}\right)\right\rangle\nn \\
= & -\frac{72hy^{2}}{c^{2}\left(z\bar{z}+y^{2}\right){}^{2}}\int_{0}^{z}dz'\int_{0}^{\overline{z}}d\overline{z}'\left(z-z'\right)^{2}\left(\overline{z}-\overline{z}'\right)^{2}\left\langle [T\left(z'\right)\overline{T}\left(\overline{z}'\right)]T\left(z_{1}\right)\overline{T}\left(\overline{w}_{1}\right)\right\rangle \nn\\
= & \frac{2hy^{2}z^{3}\bar{z}^{3}}{z_{1}^{3}\bar{w}_{1}^{3}\left(z-z_{1}\right)\left(\bar{w}_{1}-\bar{z}\right)\left(z\bar{z}+y^{2}\right){}^{2}}
\end{align}
So putting these two terms together, we get 
\begin{align}
 & \frac{\left\langle \phi\left(y,0,0\right)\mathcal{O}\left(z,\overline{z}\right)T\left(z_{1}\right)\overline{T}\left(\overline{w}_{1}\right)\right\rangle }{\left\langle \phi\left(y,0,0\right)\mathcal{O}\left(z,\overline{z}\right)\right\rangle }\\
= & \frac{h^{2}z^{2}\bar{z}^{2}\left(y^{2}\left(3\bar{w}_{1}-2\bar{z}\right)+\bar{w}_{1}z\bar{z}\right)\left(y^{2}\left(3z_{1}-2z\right)+z_{1}z\bar{z}\right)}{z_{1}^{3}\overline{w}_{1}^{3}\left(z_{1}-z\right){}^{2}\left(\overline{w}_{1}-\bar{z}\right)^{2}\left(z\bar{z}+y^{2}\right)^{2}}+\frac{2hy^{2}z^{3}\bar{z}^{3}}{z_{1}^{3}\bar{w}_{1}^{3}\left(z-z_{1}\right)\left(\bar{w}_{1}-\bar{z}\right)\left(z\bar{z}+y^{2}\right)^{2}}\nn
\end{align}
Sending $y\rightarrow0$, the second term vanishes, and the first
term will reduce to the boundary four-point function $\left\langle \mathcal{O}\left(0,0\right)\mathcal{O}\left(z,\overline{z}\right)T\left(z_{1}\right)\overline{T}\left(\overline{w}_{1}\right)\right\rangle =\left\langle \mathcal{O}\mathcal{O}T\right\rangle \left\langle \mathcal{O}\mathcal{O}\overline{T}\right\rangle =\frac{h^{2}z^{2}\overline{z}^{2}}{z^{2h}\overline{z}^{2h}z_{1}^{2}\overline{w}_{1}^{2}\left(z_{1}-z\right)^{2}\left(\overline{w}_{1}-\overline{z}\right)^{2}}$
as expected.

\subsubsection{$\left\langle \phi\mathcal{O}TT\right\rangle $ }
Using the regulator (\ref{eq:C10Regulator1}-\ref{eq:C10Regulator}), the kernels in the bulk-boundary OPE of $\phi\CO$ that contribute to $\left\langle \phi\mathcal{O} TT\right\rangle $ 
 are the identity, $K_{T}$, $K_{TT}$ and $K_{T}K_{T}$. So $\left\langle \phi\mathcal{O}TT\right\rangle $
is given by
\begin{align}\label{eq:PhiOTTUsingOPEBlock}
\frac{\left\langle \phi\left(y,0,0\right)\mathcal{O}\left(z,\overline{z}\right)T\left(z_{1}\right)T\left(z_{2}\right)\right\rangle }{\left\langle \phi\left(y,0,0\right)\mathcal{O}\left(z,\overline{z}\right)\right\rangle }= & \left\langle e^{K_{T}+K_{TT}+\cdots}T\left(z_{1}\right)T\left(z_{2}\right)\right\rangle  \\
= & \left\langle T\left(z_{1}\right)T\left(z_{2}\right)\right\rangle +\left\langle \left(K_{T}+K_{TT}+\frac{K_{T}K_{T}}{2}\right)T\left(z_{1}\right)T\left(z_{2}\right)\right\rangle.\nn
\end{align}
The first term is trivial and it's just $\left\langle T\left(z_{1}\right)T\left(z_{2}\right)\right\rangle =\frac{c}{2\left(z_{1}-z_{2}\right)^{4}}.$

The first two terms in the second braket give the following contribution
\begin{align}\label{eq:KTKTTContributionToPhiOTT}
 & \left\langle K_{T}T\left(z_{1}\right)T\left(z_{2}\right)\right\rangle +\left\langle K_{TT}T\left(z_{1}\right)T\left(z_{2}\right)\right\rangle \nonumber \\
 & =\frac{12h}{c}\int_{0}^{z}dz'\frac{\left(y^{2}+z'\overline{z}\right)\left(z-z'\right)}{z\bar{z}+y^{2}}\left\langle T\left(z'\right)T\left(z_{1}\right)T\left(z_{2}\right)\right\rangle \nonumber \\
 & \qquad+\int_{0}^{z}dz'\int_{0}^{z'}dz''\frac{72h\left(z-z'\right)^{2}\left(y^{2}+\overline{z}z''\right)^{2}}{c^{2}\left(z\bar{z}+y^{2}\right){}^{2}}\left\langle \left[T\left(z'\right)T\left(z''\right)\right]T\left(z_{1}\right)T\left(z_{2}\right)\right\rangle  \\
 & =\frac{2hz^{2}\left(y^{2}z\bar{z}z_{1}z_{2}\left(z\left(z_{1}+z_{2}\right)-4z_{1}z_{2}\right)-z^{2}\bar{z}^{2}z_{1}^{2}z_{2}^{2}+y^{4}\left(zz_{1}z_{2}\left(z_{1}+z_{2}\right)-3z_{1}^{2}z_{2}^{2}-z^{2}\left(z_{1}-z_{2}\right)^{2}\right)\right)}{\left(z-z_{1}\right)z_{1}^{3}z_{2}^{3}\left(z_{2}-z\right)\left(z_{2}-z_{1}\right){}^{2}\left(z\bar{z}+y^{2}\right)^{2}}\nn
\end{align}
where in the second line and third line, we used
\begin{align}
\left\langle T\left(z'\right)T\left(z_{1}\right)T\left(z_{2}\right)\right\rangle  & =\frac{c}{\left(z_{1}-z_{2}\right)^{2}\left(z_{2}-z'\right)\left(z_{1}-z'\right)^{2}},\\
\left\langle \left[T\left(z'\right)T\left(z''\right)\right]T\left(z_{1}\right)T\left(z_{2}\right)\right\rangle  & =\frac{c^{2}}{4}\left(\frac{1}{\left(z'-z_{1}\right)^{4}\left(z''-z_{2}\right)^{4}}+\frac{1}{\left(z'-z_{2}\right)^{4}\left(z''-z_{1}\right)^{4}}\right).\nn
\end{align}
Notice that there is no logarithm in the result of equation (\ref{eq:KTKTTContributionToPhiOTT}).
But if one computes $\left\langle K_{T}T\left(z_{1}\right)T\left(z_{2}\right)\right\rangle $
and $\left\langle K_{TT}T\left(z_{1}\right)T\left(z_{2}\right)\right\rangle $
separately, one can see that they both have logarithmic terms, but
they cancel out exactly!

The last term in the second bracket of equation (\ref{eq:PhiOTTUsingOPEBlock}) is 
\begin{align}
 & \left\langle \frac{K_{T}K_{T}}{2}T\left(z_{1}\right)T\left(z_{2}\right)\right\rangle \nn\\
= & \int_{0}^{z}dz'\int_{0}^{z}dz''\left(\frac{72h^{2}}{c^{2}}\frac{\left(y^{2}+z'\overline{z}\right)\left(z-z'\right)}{z\bar{z}+y^{2}}\frac{\left(y^{2}+z''\overline{z}\right)\left(z-z''\right)}{z\bar{z}+y^{2}}\right)\left\langle \left[T\left(z'\right)T\left(z''\right)\right]T\left(z_{1}\right)T\left(z_{2}\right)\right\rangle \nn\\
= & \frac{h^{2}z^{4}\left(zz_{1}\bar{z}+y^{2}\left(3z_{1}-2z\right)\right)\left(zz_{2}\bar{z}-2y^{2}z+3y^{2}z_{2}\right)}{\left(z-z_{1}\right){}^{2}z_{1}^{3}z_{2}^{3}\left(z_{2}-z\right){}^{2}}
\end{align}
which is just the contribution from $\left\langle \phi\mathcal{O}T\right\rangle \left\langle \phi\mathcal{O}T\right\rangle $. 

So putting everything together, we have
\begin{align}
 & \frac{\left\langle \phi\left(y,0,0\right)\mathcal{O}\left(z,\overline{z}\right)T\left(z_{1}\right)T\left(z_{2}\right)\right\rangle }{\left\langle \phi\left(y,0,0\right)\mathcal{O}\left(z,\overline{z}\right)\right\rangle }\nn\\
= & \frac{c}{2\left(z_{1}-z_{2}\right)^{4}}+\frac{h^{2}z^{4}\left(z_{1}z\bar{z}+y^{2}\left(3z_{1}-2z\right)\right)\left(z_{2}z\bar{z}+y^{2}\left(3z_{2}-2z\right)\right)}{z_{1}^{3}z_{2}^{3}\left(z-z_{1}\right){}^{2}\left(z-z_{2}\right)^{2}\left(z\bar{z}+y^{2}\right)^{2}}\\
 & +\frac{2hz^{2}\left(y^{2}z\bar{z}z_{1}z_{2}\left(z\left(z_{1}+z_{2}\right)-4z_{1}z_{2}\right)-z^{2}\bar{z}^{2}z_{1}^{2}z_{2}^{2}+y^{4}\left(zz_{1}z_{2}\left(z_{1}+z_{2}\right)-3z_{1}^{2}z_{2}^{2}-z^{2}\left(z_{1}-z_{2}\right)^{2}\right)\right)}{\left(z-z_{1}\right)z_{1}^{3}z_{2}^{3}\left(z_{2}-z\right)\left(z_{2}-z_{1}\right){}^{2}\left(z\bar{z}+y^{2}\right)^{2}}\nn
\end{align}
Sending $y\rightarrow0$ the above result does give us $\frac{\left\langle \mathcal{O}\left(0,0\right)\mathcal{O}\left(z,\overline{z}\right)T\left(z_{1}\right)T\left(z_{2}\right)\right\rangle }{\left\langle \mathcal{O}\left(0,0\right)\mathcal{O}\left(z,\overline{z}\right)\right\rangle }$,
which is 
\begin{align}
\frac{\left\langle \mathcal{O}\left(0,0\right)\mathcal{O}\left(z,\overline{z}\right)T\left(z_{1}\right)T\left(z_{2}\right)\right\rangle }{\left\langle \mathcal{O}\left(0,0\right)\mathcal{O}\left(z,\overline{z}\right)\right\rangle } & =\frac{1}{\left(z_{1}-z_{2}\right)^{4}}\left[\frac{c}{2}+\frac{hu^{2}(u(hu-2)+2)}{(u-1)^{2}}\right]\\
 & =\frac{c}{2\left(z_{1}-z_{2}\right){}^{4}}+\frac{hz^{2}\left(hz^{2}+\frac{2z_{1}z_{2}\left(z-z_{1}\right)\left(z-z_{2}\right)}{\left(z_{1}-z_{2}\right){}^{2}}\right)}{z_{1}^{2}z_{2}^{2}\left(z-z_{1}\right){}^{2}\left(z-z_{2}\right)^{2}}\nn
\end{align}
where $u\equiv\frac{z_{12}z_{34}}{z_{13}z_{24}}=\frac{\left(z_{1}-z_{2}\right)z}{\left(z_{1}-z\right)z_{2}}$
is the cross ratio.

\subsection{Spinning Bulk Wilson Lines}
\label{app:SpinningBulkWL}

In this appendix we give the derivation of equation (\ref{eq:spinningBulkWL}) in the text.  To begin, we recall how to write the bulk-to-boundary propagators in the vacuum.  The general procedure was described in \cite{Katz}, and takes the form
\be
\< A^{\mu_1, \dots, \mu_\ell}(y, z_1, \bar{z}_1) \CO_{h,\bar{h}}(z_2, \bar{z}_2) \> = \left( \frac{y}{y^2+ z_{12} \bar{z}_{12}} \right)^{2h} \xi_\pm^{\mu_1} \dots \xi_\pm ^{\mu_\ell} , \quad (\pm = -{\rm sgn}(\ell)),
\ee
for the case $h-\bar{h}=\ell$ of interest.  
Here, $(\xi_+^y, \xi_+^z, \xi_+^{\bar{z}} ) = (yz_1, z_1^2, -y^2)$ is the Killing vector associated with holomorphic special conformal generators, and $(\xi_-^y, \xi_-^z, \xi_-^{\bar{z}}) = (y \bar{z}_1, -y^2, \bar{z}_1^2)$  for anti-holomorphic ones.  

To promote this to an arbitrary background, we perform the transformation (\ref{eq:GeneralVacDiffeomorphism}).  Because $A_{\mu_1, \dots, \mu_\ell}$ is a tensor, this transformation includes factors of
\be
\frac{\partial x_f^\mu}{\partial x^{\mu'}} ,
\ee
where $x_f$ are the transformed coordinates $(y_f, z_f, \bar{z}_f)$.  The transformed coordinates include dependence on the second derivatives of $f,\bar{f}$, and so the above Jacobian factor depends on its third derivatives.  These third derivatives $f'''(z_2), \bar{f}'''(\bar{z}_2)$ can be eliminated in terms of the stress tensor $T(z_2), \bar{T}(\bar{z}_2)$ at the point $(z_2, \bar{z}_2)$.  Moreover, as before we can eliminate $f'(z_2),f''(z_2)$ in terms of $x_T(z_1)$ and $E_T$.  Making such substitutions, we find that 
\be
(\xi'_-)_{\mu'} = \frac{\partial x^\mu}{\partial x_f^{\mu'}} (\xi_-)_{\mu}(u_2, w_2, \bar{w}_2) =  t^\mu_{\mu'} \bar{E}_{\bar{T}}^{-2} \bar{f}'(\bar{z}_1)  (\xi_-)_\mu(y,z,\bar{z}) ,
\label{eq:ximinustfmd}
\ee
where $t^\mu_{\mu'}$ is given in (\ref{eq:spinningextrapiece}).     We also have, from massaging (\ref{eq:BBPtfmed}) a bit, that
\be
 \left( \frac{u_2}{u_2^2 + f_{21}\bar{f}_{21}} \right)^{2h} =  E_T^{2h} \bar{E}_{\bar{T}}^{2h}  (f'(z_1) \bar{f}'(\bar{z}_1))^{-h} \left( \frac{y}{y^2 + x_T(z_1) \bar{x}_{\bar{T}}(\bar{z}_1)} \right)^{2h} . 
\ee
Multiplying by $(\xi'_-)_{\mu'_1} \dots (\xi'_-)_{\mu'_\ell}$ and using (\ref{eq:ximinustfmd}), we find
\be
\left( \frac{u_2}{u_2^2 + f_{21}\bar{f}_{21}} \right)^{2h} (\xi'_-)_{\mu'_1} \dots (\xi'_-)_{\mu'_\ell} &=& E_T^{2h} \bar{E}_{\bar{T}}^{2\bar{h}} (f'(z_1))^{-h} (f'(\bar{z}_1))^{-\bar{h}} \\
 & \times &\left( \frac{y}{y^2 + x_T(z_1) \bar{x}_{\bar{T}}(\bar{z}_1)} \right)^{2h} t^{\mu'_1}_{\mu_1} \dots t^{\mu'_\ell}_{\mu_\ell} (\xi_-)_{\mu_1} \dots (\xi_-)_{\mu_\ell}. \nn
\ee
Equation (\ref{eq:spinningBulkWL}) follows by using the fact that the Wilson line factors simply impose the constraint $x\rightarrow x_T(z_1), \bar{x} \rightarrow x_{\bar{T}}(\bar{z}_1)$ and produce factors $E_T^{2h}, \bar{E}_{\bar{T}}^{2\bar{h}}$.

\subsection{Bulk Witten Diagram Computation for $\<\phi\CO T\>$}
\label{app:BulkWittenDiagramComputation}

\begin{figure}[h]
\begin{center}
\includegraphics[width=.4\textwidth]{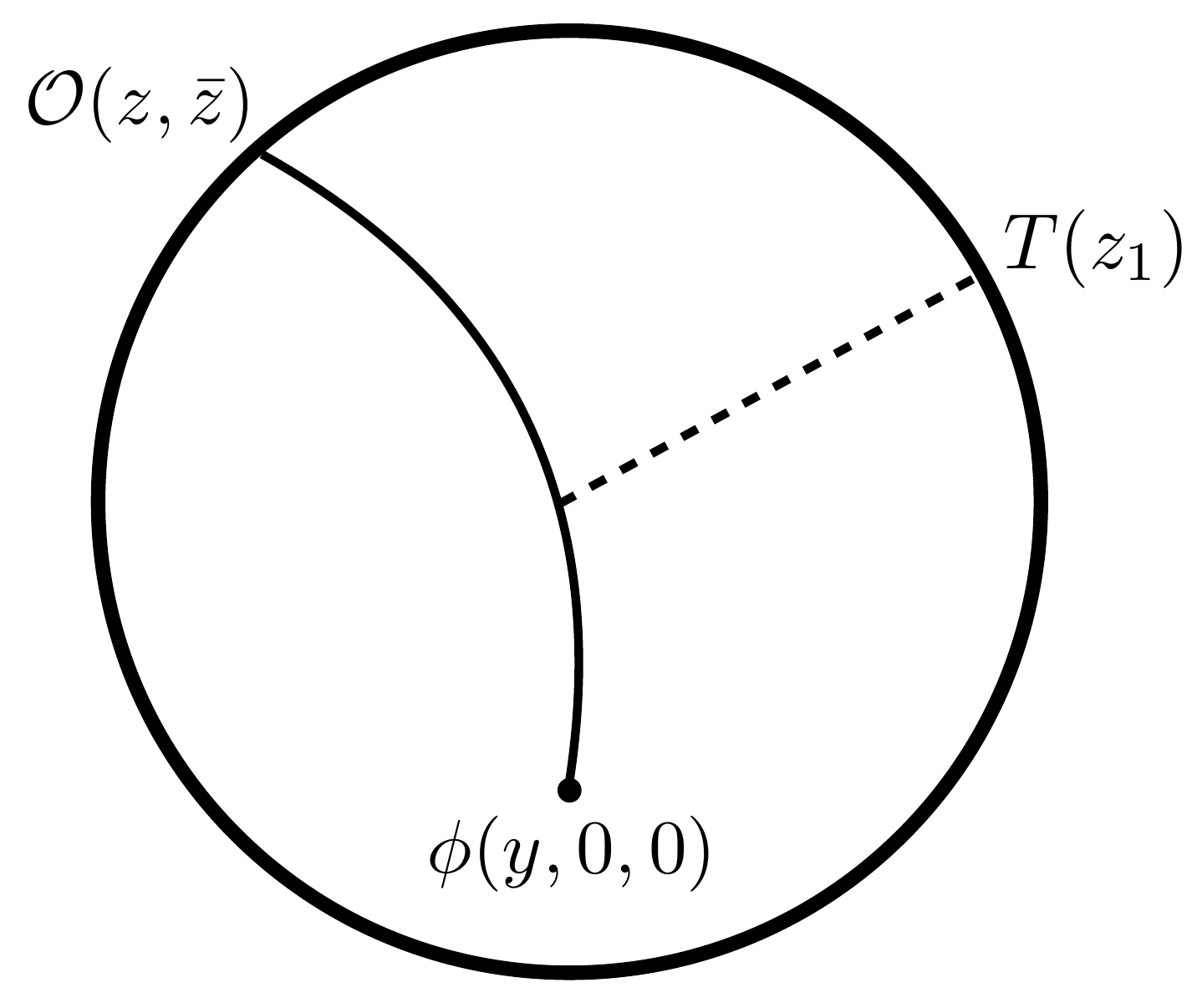}
\label{fig:PhiOTSetup} 
\caption{Dashed (solid) lines are graviton (scalar) propagators.}
\end{center}
\end{figure}

In this section, we will show that the result we obtained for $\<\phi\CO T\>$ using bulk-boundary OPE block and the recursion relation agrees with the result of the bulk Witten diagram computation for $\langle \phi(y,0,0)\mathcal{O}(z,\bar{z}) T(z_1)\rangle$, shown in Fig. \ref{fig:PhiOTSetup}.  This should be expected, as the definition of equation (\ref{eq:OPEBlockAsPathIntegral}) is essentially the first-quantized version of the bulk field theory that leads to the Witten diagram we will discuss.  We will first show that the result is exact, using a trick  \cite{Howtozintegrals} that obviates the need to perform integrals over AdS$_3$. Then we will  explicitly evaluate the diagram in the large $h$ limit using saddle point approximation (this will give the same exact result), where we can make direct contact with some of the results from section \ref{sec:GravitationalWilsonLines}.  

In order to compute this diagram, we need four ingredients: the scalar bulk-to-boundary propagator, the scalar bulk-to-bulk propagator,  the vertex structure associated with the scalar-graviton interaction, and the graviton bulk-to-boundary propagtor. The standard prescription is to  multiply these propagators together, and integrate over the bulk. There are a variety of conventions for normalizing these objects, so we will mostly ignore the overall numerical prefactors, which can be fixed in any case in terms of operator normalizations and the stress tensor Ward identity.

The bulk-to-bulk propagator, specializing to our coordinate set-up, is given by
\be
G_{(y,0,0),(y',z',\bar{z}')} =  \frac{e^{-2h \sigma}}{1-e^{-2\sigma}},
\ee
where $\sigma\equiv\sigma_{(y',z',\bar{z}'),(y,0,0)}$ is the bulk-bulk geodesic between $(y',z',\bar{z}')$ and $(y,0,0)$ 
\begin{equation}
	\sigma_{(y',z',\bar{z}'),(y,0,0)}= \log \frac{1+\sqrt{1-\xi^2}}{\xi}, \quad\quad\text{with } \quad\xi = \frac{2y y'}{y^2 + y'^2 + z' \bar{z}'}.
\end{equation}
The scalar bulk-to-boundary propagator is given by \begin{equation}
	K_{(y',z',\bar{z}'),(z,\bar{z})} =  \left(\frac{y'}{y'^2 + (z-z')(\bar{z}-\bar{z}')} \right)^{2h},
\end{equation}
 while can also be written as 
 \begin{equation}
	K_{(y',z',\bar{z}'),(z,\bar{z})}= e^{-2h\sigma_{(y',z',\bar{z}'),(z,\bar z)}},
\end{equation}
where $\sigma_{(y',z',\bar{z}'),(z,\bar z)}=\log\frac{y'^2+ (z'-z) (\bar{z}'-z)}{y'}$ is the regulated bulk-boundary geodesic length.

The vertex structure is given by $h_{\mu\nu}T_m^{\mu\nu}$, where $T_m^{\mu\nu}$ is the bulk matter stress energy tensor. It can be derived from the bulk equations of motion, and is given by \cite{D'Hoker} \begin{equation}
	T_m^{\mu\nu} = (g^{\mu\alpha} g^{\nu\beta} + g^{\mu\beta} g^{\nu\alpha}) \partial_\alpha K \partial_\beta G - g^{\mu\nu} (g^{\rho\alpha} \partial_\rho K \partial_\alpha G + m^2 K G).
\end{equation} We are interested in the holomorphic part of this tensor object, since the coupling we need is $h_{zz} T_m^{zz}$. In the Fefferman-Graham gauge, it simplifies to \begin{equation}
	T_m^{zz} = 2 g^{z\bar{z}} g^{z\bar{z}} \partial_{\bar{z}} K \partial_{\bar{z}} G = - 2 y^4 G \partial_{\bar{z}}^2 K.
\end{equation} Finally, we need the graviton bulk-to-boundary propagator in this gauge. $h_{zz}(y,z,\bar z)$ is by definition equal to $-\frac{6T(z)}{c}$, as in equation (\ref{eq:MetricwithT}).\footnote{It was also shown \cite{Kabat:2012hp} using smearing functions that $h_{zz}(y,z,\bar z)$ is simply given by boundary stress energy tensor $T(z)$:
 \begin{align}
h_{zz}\left(y,z,\overline{z}\right) & \propto\frac{1}{\pi y^{2}}\int_{z\overline{z}\le y^{2}}dz'\overline{z}'T_{zz}\left(z+iz'\right)\nn\\
 & =\frac{1}{\pi y^{2}}\int_{0}^{y}rdr\int_{0}^{2\pi}d\theta T\left(z+ire^{i\theta}\right)\\
 & =T\left(z\right).\nn
\end{align}}
So we have 
\begin{equation}
	\<h_{zz}(y',z',\bar z')T(z_1)\>=-\frac{6}{c}\<T(z')T(z_1)\>=\frac{-3}{(z'-z_1)^4}
\end{equation}
Putting these ingredients together, the bulk integral corresponding to fig. \ref{fig:PhiOTSetup} is then 
\begin{equation} \label{eq:WittenIntegral}
\begin{aligned}
	\langle \phi(y,z_3,\bar{z}_3)\mathcal{O}(z_2,\bar{z}_2)T(z_1)\rangle &= \int_{\textrm{AdS}_3} \sqrt{g} dz' d\bar{z}' dy' (-2 y'^4) \\
	&\times G_{(y,z_3,z_3),(y',z',\bar{z}')} \partial_{\bar{z}}^2 K_{(y',z',\bar{z}'),(z_2,\bar{z}_2)} \frac{-3}{(z'-z_1)^4}. \end{aligned}
\end{equation} 

The trick  \cite{Howtozintegrals} to evaluating this kind of Witten diagram integral is first to simplify the problem as much as possible using global conformal invariance, and second to recall that the bulk scalar Feynman propagator satisfies the Klein-Gordon equation
\be
\left(\nabla^2 -  m^2 \right) G(X,Y) = \delta_{AdS}(X-Y),
\ee
where $m^2 = 2h(2h-2)$.  This means that if we act with the bulk differential operator $(\nabla^2 -  m^2)$ on the Witten diagram that computes $\< \phi(X) \CO(z_2) T(z_1) \>$, then we will be left with just the integrand above, with $G$ removed.  We can simplify the calculation by shifting $z_2$ to 0 with a translation, then performing an inversion, and finally shifting $z_3 \rightarrow 0$ by another translation.\footnote{ Because of the presence of the bulk coordinate $y$, it is not enough to just take $z_2 \rightarrow \infty$, rather, we must actually perform the transformation ($z\rightarrow z-z_2$ followed by an inversion) that takes $z_2 \rightarrow \infty$.  }  The resulting equation of motion is
\be
(\nabla^2 - m^2) A(y, z_3, \bar{z}_3) &=&-12 \Delta(\Delta+1) \bar{z}_1^4  y^\Delta \left( \frac{y}{y^2 + z_3 \bar{z}_3} \right)^4,  \nn\\
A(y,z_3, \bar{z}_3) &\equiv & \langle \phi(y',z'_3,\bar{z}'_3)\mathcal{O}(z'_2, \bar{z}'_2)T(z'_1)\rangle,
\ee
where $(y',z_i')$ are the transformed coordinates.  For comparison, the result in (\ref{eq:PhiOT}) in terms of the transformed coordinates is
\be
\langle \phi(y',z'_3,\bar{z}'_3)\mathcal{O}(z'_2, \bar{z}'_2)T(z'_1)\rangle &=& \frac{\Delta}{2} \frac{y^\Delta \bar{z}^4_1 z_3^2 (3 y^2 + z_3 \bar{z}_3)}{(y^2 + z_3 \bar{z}_3)^3} = \frac{\Delta}{2} y^{\Delta-4} z^2_3 t^2 (1+2t), \nn\\
\ee
where $t \equiv \frac{y^2}{y^2+ z_3 \bar{z}_3}.$  
Taking $A(y, z_3, \bar{z}_3) = y^{\Delta-4} z^2 f(t)$, the equation of motion is simply
\be
f''(t)+\frac{(-\Delta +(\Delta -1) t+4) f'(t)}{(t-1) t}+\frac{2 (\Delta -3) f(t)}{(t-1) t^2}-\frac{3 \Delta  (\Delta +1) t^2}{t-1} &=& 0.
\ee
It is straightforward to check that the result in (\ref{eq:PhiOT}), i.e. $f(t) = \frac{\Delta}{2} t^2 (1+2t)$, satisfies this equation.  More constructively, there are two boundary conditions that must be imposed to fix the solution; one of these is that there is no $y^{2-\Delta}$ piece near the boundary, and the other can be chosen so that the correct $\< \CO \CO T\>$ three-point function is reproduced at $y \sim 0$; since (\ref{eq:PhiOT}) manifestly satisfies these conditions, it is the correct solution. Thus our result exactly matches the Witten diagram.

Next, at large $h$, we can also evaluate the integral  (\ref{eq:WittenIntegral}) directly using saddle point approximation (the result of this saddle point approximation turns out to be exact) and see how the kernel (\ref{eq:KT}) emerges.  After some manipulations, the bulk integral (\ref{eq:WittenIntegral}) can be re-cast into a more suggestive form 
\begin{small}
\begin{equation}
	\langle\phi(y,0,0)\mathcal{O}(z,\bar{z})T(z_1)\rangle = 12h(2h+1) \int_{\textrm{AdS}_3}  \frac{dz' d\bar{z}' dy'}{y'^3} e^{-2h L(y',z',\bar{z}')} \frac{e^{-2\sigma_{(y',z',\bar{z}'),(z,\bar{z})}}}{1-e^{-2\sigma_{(y,0,0),(y',z',\bar{z}')}}} y'^2\frac{(z'-z)^2}{(z'-z_1)^4}. \label{bulkint}
\end{equation} 
\end{small}
The notation $\sigma_{a,b}$ indicates the (regulated) geodesic length between points $a$ and $b$. We have also defined $ L(y',z',\bar{z}') $ to be the sum of the lengths of geodesics from $(y,0,0)$ to $(y',z',\bar z')$ and from $(y',z',\bar z')$ to $(z,\bar z)$, that is
 \begin{equation} L(y',z',\bar{z}') \equiv \sigma_{(y,0,0),(y',z',\bar{z}')} + \sigma_{(y',z',\bar{z}'),(z,\bar{z})}.
\end{equation} 
In the large $h$ limit, the integral will localize  along the geodesics from $(y,0,0)$ to $(z,\bar z)$ to minimize $L$. This geodesic parameterized by $z'$ is given by
 \begin{equation}\label{eq:Geodesic}
\bar{z}' = \frac{\bar{z}}{z}z', \quad\quad\quad  y'^2 = \left(1-\frac{z'}{z} \right)(y^2 + z'\bar{z}), 
\end{equation} 
 so that the saddle point approximation to equation (\ref{bulkint}) is 
\begin{equation}
	\begin{aligned}
		\langle \phi\mathcal{O}T\rangle &\propto \frac{24 h^2}{c}  e^{-2h L(y,0,0)} \int_0^z dz' \frac{1}{\sqrt{\det \partial^2 L} } \frac{e^{-2\sigma_{(y',z',\bar{z}'),(z,\bar{z})}}}{1- e^{-2\sigma_{(y,0,0),(y',z',\bar{z}')}}} \frac{1}{y'}\frac{(z-z')^2}{(z'-z_1)^4},
	\end{aligned}
\end{equation} 
where the determinant is given by \begin{equation}\det \partial^2 L = \det \begin{pmatrix} \partial_{\bar{z}'}^2 L & \partial_{\bar{z}'} \partial_{y'} L \\
\partial_{y'} \partial_{\bar{z}'} L  & \partial_{y'}^2 L\end{pmatrix} = \frac{4 z^5 (z'\bar{z}+y^2)}{z'^2(z'-z)(z\bar{z}+y^2)^4},	
\end{equation} 
evaluated along the geodesic (\ref{eq:Geodesic}). Plugging this in (and neglecting an order 1 numerical factor) and performing the $z'$ integral, we obtain \begin{align}
	\langle \phi(y,0,0)\mathcal{O}(z,\bar{z})T(z_1)\rangle \propto& \frac{12 h}{c} \langle \phi(y,0,0) \mathcal{O}(z,\bar{z}) \rangle \int_0^z dz' \frac{2(z-z')(z'\bar{z}+y^2)}{z \bar{z}+y^2} \frac{c}{2(z'-z_1)^4}\nn\\
	=&\langle \phi(y,0,0) \mathcal{O}(z,\bar{z}) \rangle\frac{hz^2 }{ z_1^3
   \left( z_1 - z \right)^2 } \left( z_1+ \frac{2 y^2(z_1-z) }{y^2+z \bar z}   \right)
	\end{align}
matching equation \ref{eq:PhiOT} as expected.  This demonstrates how the kernel of equation (\ref{eq:KT}) emerges from a bulk Witten diagram calculation. 

\subsection{Solving for the Quantum Operator $\phi$}
\subsubsection{Solutions to the Conditions of equation (\ref{eq:Conditions}) at Level 3 and Level 4 }\label{app:Level3Level4Solution}
In this section, we provide the solutions to the conditions of equation (\ref{eq:Conditions}) at level 3 and level 4.

At level 3, $\left|\phi\right\rangle _{3}=\lambda_{3}\mathcal{L}_{-3}\overline{\mathcal{L}}_{-3}\left|\mathcal{O}\right\rangle $ and $\lambda_{3}\mathcal{L}_{-3}$ is given by
with
\be
\lambda_{3}\mathcal{L}_{-3}=\left(-1\right)^{3}\left(\frac{L_{-1}^{3}}{\left|L_{-1}\mathcal{O}\right|^{2}}+\frac{L_{-1}\mathcal{L}_{-2}^{\text{quasi}}}{\left|L_{-1}\mathcal{L}_{-2}^{\text{quasi}}\mathcal{O}\right|^{2}}+\frac{\mathcal{L}_{-3}^{\text{quasi}}}{\left|\mathcal{L}_{-3}^{\text{quasi}}\right|^{2}}\right),
\ee
where $\mathcal{L}_{-3}^{\text{quasi}}=L_{-1}^{3}-2\left(h+1\right)L_{-1}L_{-2}+\left(h+1\right)\left(h+2\right)L_{-3}$
and the norms are 
\begin{align*}
\left|L_{-1}\mathcal{L}_{-2}^{\text{quasi}}\mathcal{O}\right|^{2} & =2\left(h+2\right)\left|\mathcal{L}_{-2}^{\text{quasi}}\mathcal{O}\right|^{2}=\frac{4\left(2h+1\right)\left(h+2\right)\left(\left(2h+1\right)c+2h\left(8h-5\right)\right)}{9},\\
\left|\mathcal{L}_{-3}^{\text{quasi}}\right|^{2} & =2h\left(h+1\right)\left(h+2\right)\left(\left(c-7\right)h+c+3h^{2}+2\right).
\end{align*}

At level 4, $\left|\phi\right\rangle _{4}=\lambda_{4}\mathcal{L}_{-4}\overline{\mathcal{L}}_{-4}\left|\mathcal{O}\right\rangle$ and $\lambda_{4}\mathcal{L}_{-4}$ is given by 
\be
\lambda_{4}\mathcal{L}_{-4}=\frac{L_{-1}^{2}\mathcal{L}_{-2}^{\text{quasi}}}{\left|L_{-1}^{2}\mathcal{L}_{-2}^{\text{quasi}}\right|^{2}}+\frac{L_{-1}\mathcal{L}_{-3}^{\text{quasi}}}{\left|L_{-1}\mathcal{L}_{-3}^{\text{quasi}}\right|^{2}}+b_{4,1}\mathcal{L}_{-4}^{\text{quasi,\ensuremath{\left(4,1\right)}}}+b_{2,2}\mathcal{L}_{-4}^{\text{quasi,\ensuremath{\left(2,2\right)}}}
\ee
where 
\begin{footnotesize}
\begin{align*}
\mathcal{L}_{-4}^{\text{quasi,\ensuremath{\left(4,1\right)}}} & =L_{-1}^{4}-\frac{4(2h+3)}{375}\left[\left(16h\left(2h+11\right)+267\right)L_{-4}-5\left(6h+9\right)L_{-2}^{2}-5\left(16h+49\right)L_{-1}L_{-3}+125L_{-1}^{2}L_{-2}\right],\\
\mathcal{L}_{-4}^{\text{quasi,\ensuremath{\left(2,2\right)}}} & =L_{-1}^{4}+\frac{16}{9}h(h+3)L_{-2}^{2}+\left(\frac{8h}{3}+10\right)L_{-1}L_{-3}-\frac{4}{3}(2h+3)L_{-1}^{2}L_{-2}+-4(h+3)L_{-4}.
\end{align*}
\end{footnotesize}
$\mathcal{L}_{-4}^{\text{quasi,\ensuremath{\left(4,1\right)}}}$ and
$\mathcal{L}_{-4}^{\text{quasi,\ensuremath{\left(2,2\right)}}}$ are
not orthogonal to each other. $\mathcal{L}_{-4}^{\text{quasi,\ensuremath{\left(4,1\right)}}}$ becomes
a null-state when $c=c_{4,1}\left(h\right)=-\frac{8h}{5}-\frac{45}{2h+3}+\frac{53}{5}$,
and $\mathcal{L}_{-4}^{\text{quasi,\ensuremath{\left(2,2\right)}}}$
becomes a null-state when $c=c_{2,2}\left(h\right)=1-8h$. The coefficients of them,  $b_{4,1}$ and $b_{2,2}$ are given by
\begin{align*}
b_{4,1} & =\frac{1125(10c+116h-81)}{8(2h+3)(2h+5)(8h-3)(8h+27)(5c(2h+3)+2(h-1)(8h-33))(2ch+c+2h(8h-5))},\\
b_{2,2} & =\frac{81(2h(16h+19)-5c)}{16h(h+3)(2h+5)(8h-3)(8h+27)(c+8h-1)(2ch+c+2h(8h-5))}.
\end{align*}
They are actually the solution to 
\[
\left(\begin{array}{cc}
\left\langle \mathcal{O}\left|\left(\mathcal{L}_{-4}^{\text{quasi,\ensuremath{\left(4,1\right)}}}\right)^{\dagger}\mathcal{L}_{-4}^{\text{quasi,\ensuremath{\left(4,1\right)}}}\right|\mathcal{O}\right\rangle  & \left\langle \mathcal{O}\left|\left(\mathcal{L}_{-4}^{\text{quasi,\ensuremath{\left(4,1\right)}}}\right)^{\dagger}\mathcal{L}_{-4}^{\text{quasi,\ensuremath{\left(2,2\right)}}}\right|\mathcal{O}\right\rangle \\
\left\langle \mathcal{O}\left|\left(\mathcal{L}_{-4}^{\text{quasi,\ensuremath{\left(2,2\right)}}}\right)^{\dagger}\mathcal{L}_{-4}^{\text{quasi,\ensuremath{\left(4,1\right)}}}\right|\mathcal{O}\right\rangle  & \left\langle \mathcal{O}\left|\left(\mathcal{L}_{-4}^{\text{quasi,\ensuremath{\left(2,2\right)}}}\right)^{\dagger}\mathcal{L}_{-4}^{\text{quasi,\ensuremath{\left(2,2\right)}}}\right|\mathcal{O}\right\rangle 
\end{array}\right)\left(\begin{array}{c}
b_{4,1}\\
b_{2,2}
\end{array}\right)=\left(\begin{array}{c}
1\\
1
\end{array}\right).
\]
One can show that for non-orthogonal quasi-primaries at higher order,
their coefficients will be given by the solutions to the
equation corresponding to the above one at that order. And for global descendants of these non-orthogonal
quasi-primaries, their coefficients will be given by a similar equation. These equations can be derived using the method similar to the one in section \ref{sec:SolutionQP}.

\label{sec:GettingPhiNPerturbatively}
\subsubsection{From Vacuum Sector Correlators to $\phi$ Via the OPE}
We determined the vacuum sector correlators 
\be
\< \phi(X) \CO(z) T(z_1) \cdots T(z_n) \bar T(\bar z_1) \cdots \bar T(\bar z_m) \> 
\ee
using the bulk-boundary OPE block in section \ref{sec:GravitationalWilsonLines}.  Thus  we can straightforwardly determine the BOE expansion, expressing $\phi_N$ in terms of Virasoro descendants of $\CO$ by studying the multi-OPEs of $\CO$ with the various stress tensors.  

To perform this analysis explicitly, we start with the $\< \phi \CO \>$ correlator and then add more and more $T$ and $\bar T$, modifying $\phi_N$ each time to obtain the correct correaltors.  We already found that global BOE of equation (\ref{eq:PhiGlobalBOE}) produces the correct $\< \phi \CO \>$ correlator (see appendix \ref{app:GlobalBOEBasics} for details).  Thus the next step is to modify the BOE to achieve the correct $\< \phi \CO T \>$ correlators, without disrupting $\< \phi \CO \>$.  For this purpose it is useful to compute
\be
\< \phi^{\text{global}}(y,0,0) \CO(z) T(z_1) \> =  \frac{h \left(z \bar z +y^2 \right)^2}{\left(z_1-z \right)^2 \left(z_1 \bar z + y^2 \right)^{2}  } \, \<\phi(y) \CO(z) \>
\ee
as shown via a more general argument in appendix \ref{app:GlobalBOEBasics}.  Now we can subtract this result from the full correlator in equation (\ref{eq:PhiOT}) to obtain correlators of $\phi_N$ with the contributions of global conformal descendants of $\CO$ removed.  Expanding to low order in $y$, this is
\be
\label{eq:SeriesVirMinusGlobal}
\left( \frac{z \bar z}{y} \right)^h \left\<  \CO(z) T(z_1) \left( \phi - \phi^{\text{global}} \right) \right\> 
= -\frac{3 h y^4}{z_1^4 \bar z^2} +
2 h y^6 \left(\frac{1}{z_1^4 z \bar z^3}+\frac{2}{z_1^5 \bar z^3}\right) + \cdots
\ee
Notice that the expansion only begins at order $y^4$, and that as a function of $z_1$, the location of the stress tensor, each term has a pole at the origin of order $4$ or higher.  The first observation indicates that the first Virasoro correction occurs in $\phi_2$, while the second confirms that these corrections all involve Virasoro descendants of $\CO$, ie new quasi-primaries like $[T \CO]$.  We can match to the Virasoro descendants at levels $2$ and $3$, namely the operators $L_{-2} \bar L_{-1}^2 \CO$, $L_{-3} \bar L_{-1}^3 \CO$, and $L_{-1} L_{-2} \bar L_{-1}^3 \CO$, by computing correlators such as
\be
\<  \CO(z, \bar z) T(z_1) L_{-2} \bar L_{-1}^2 \CO(0) \> \approx \frac{2h(2h+1)}{\bar z^{2h+2} z^{2h}}  \frac{c}{2 z_1^4} 
\ee
where we have neglected terms that are independent of $c$.  Comparing this with equation \ref{eq:SeriesVirMinusGlobal} at large $c$, we see that we need to add
\be
\delta \phi_2 \approx -\left( \frac{y^4}{2! (2h)_2 } \right) \frac{12h}{c} L_{-2} \bar L_{-1}^2 \CO(0)
\ee
to $\phi_2$ at this order.   At order $y^6$ we would add a linear combination of $L_{-3} \bar L_{-1}^3 \CO$ and $L_{-1} L_{-2} \bar L_{-1}^3 \CO$.  

The second step in the analysis is to go back and `fix' the $\< \phi \CO \>$ correlators, as $\delta \phi_2$ above will alter it.  To achieve this goal, we simply need to supplement $\delta \phi_2$ to make it proportional to a new level 2 quasi-primary.  This leads to
\be
\delta \phi_2 \approx \left( \frac{ y^4}{2! (2h)_2 } \right) \left(L_{-1}^2-\frac{12h}{c}L_{-2}\right) \bar L_{-1}^2 \CO(0)
\ee
to leading order at large $c$.
With this choice, $\delta \phi_2$  will have a vanishing correlator with $\CO$, and thus $\< \phi \CO \>$ will remain correct.

However, we can determine all of these coefficents more precisely and systematically using the condition of equation (\ref{eq:Conditions}), as we'll do in next subsection.

\subsubsection{Solving for $\phi$ at Large $c$}
\label{app:SolutionAtLargec}

In this section, we'll use the definition of $\phi$ to derive the leading order terms of the $\frac{1}{c}$ and $\frac{1}{c^{2}}$ corrections to $\phi$.

We know that at the leading order of the large $c$ limit, $\phi\left(y,0,0\right)$
will reduce to  $\phi^{\text{global}}\left(y,0,0\right)$, that
is 
\begin{equation}
\lim_{c\rightarrow\infty}\phi\left(y,0,0\right)\left|0\right\rangle =\phi^{\text{global}}\left(y,0,0\right)\left|0\right\rangle =\sum_{N=0}^{\infty}y^{2h+2N}\frac{\left(-1\right)^{N}}{N!\left(2h\right)_{N}}\left(L_{-1}\overline{L}_{-1}\right)^{N}\left|\mathcal{O}\right\rangle .
\end{equation}
 We'll expand $\phi(y,0,0)|0\>=\sum_{N=0}^\infty y^{2h+2N}|\phi\>_N$ and write
$\left|\phi\right\rangle _{N}$ as follows
\begin{equation}
\left|\phi\right\rangle _{N}=\lambda_{N}\mathcal{L}_{-N}\overline{\mathcal{L}}_{-N}\left|\mathcal{O}\right\rangle .
\end{equation}
And we'll derive the coefficients of the following terms at order
$\frac{1}{c}$ and $\frac{1}{c^{2}}$ in $\mathcal{L}_{-N}$:
\begin{equation}\label{eq:1OvercSquareInMathematicalL}
\mathcal{L}_{-N}=L_{-1}^{N}+\frac{1}{c}\sum_{k=2}^{N}\eta_{N,k}L_{-k}L_{-1}^{N-k}+\frac{1}{c^{2}}\sum_{\substack{k_{1},k_{2}=2\\
k_{1}\ge k_{2}
}
}^{N}\kappa_{N,k_{1},k_{2}}L_{-k_{1}}L_{-k_{2}}L_{-1}^{N-k_{1}-k_{2}}+\CO(c^{-3}).
\end{equation}

To derive $\eta_{N,k}$, we just need to consider the first two terms in the above equation. Using the condition of equation (\ref{eq:Conditions}) we have
\begin{align}
L_{m}\left[L_{-1}^{N}+\frac{1}{c}\sum_{k=2}^{N}\eta_{N,k}L_{-k}L_{-1}^{N-k}\right]\left|\mathcal{O}\right\rangle  & =0+\mathcal{O}\left(c^{-1}\right),\qquad2\le m\le N.\label{eq:Lmequal0}
\end{align}
The first term can be calculated exactly as follows\footnote{Equation (\ref{eq:LmLMinus1N}) comes from the following procedure. We commute $L_{m}$
with $m$ $L_{-1}$ to get $L_0$. To do so we need to choose $m$ $L_{-1}$s
from the $N$ $L_{-1}$s. If the position of the last $L_{-1}$ for these
$m$ $L_{-1}$s is the $i$th $L_{-1}$ in the $N$ $L_{-1}$s from
the right, then it means that we need to choose $\left(m-1\right)$
$L_{-1}$s from $\left(n-i\right)$ $L_{-1}$s, where there are $\left(\begin{array}{c}
N-i\\
m-1
\end{array}\right)$ of ways to do so. Commuting $L_{m}$ with $m$ $L_{-1}$ will eventually
gives us a $L_{0}$ times a factor of $\left(m+1\right)!$. And there are $(i-1)$ $L_{-1}$s remained on the right of this $L_0$, so the eigenvalue
of $L_{0}$ will be $h+i-1$.}
\begin{align}\label{eq:LmLMinus1N}
L_{m}L_{-1}^{N}\left|\mathcal{O}\right\rangle  & =\left(m+1\right)!\sum_{i=1}^{N-\left(m-1\right)}\left(\begin{array}{c}
N-i\\
m-1
\end{array}\right)\left(h+i-1\right)L_{-1}^{N-m}\left|\mathcal{O}\right\rangle \nn\\
 & =\frac{N!(h\left(m+1\right)+N-m)}{\left(N-m\right)!}L_{-1}^{N-m}\left|\mathcal{O}\right\rangle 
\end{align}
The second term is easy to calculate at leading order of large $c$,
which is given by 
\begin{equation}\label{eq:LmLMinusk}
L_{m}\sum_{k=2}^{N}\frac{1}{c}\eta_{N,k}L_{-k}L_{-1}^{N-k}\left|\mathcal{O}\right\rangle =\eta_{N,m}\frac{m\left(m^{2}-1\right)}{12}L_{-1}^{N-m}\left|\mathcal{O}\right\rangle +\CO(c^{-1})
\end{equation}
where we used the Virasoro algebra $[L_m,L_n]=(m-n)L_{m+n}+ \frac{m(m^2-1)c}{12}\delta_{m,-n}$.Equating the RHSs of  equation (\ref{eq:LmLMinus1N}) and equation (\ref{eq:LmLMinusk}), and  solving for $\eta_{N,m}$,
we find
\be\label{eq:EtaNm}
\eta_{N,m}  =-\frac{12(h\left(m+1\right)+N-m)N!}{\left(N-m\right)!m\left(m^{2}-1\right)}
\ee
To derive $\kappa_{N,k_{1},k_{2}}$, we need to use the following conditions, 
\begin{footnotesize}
\begin{equation}\label{eq:1OvercSquareCorrectionToPhi}
L_{m_{2}}L_{m_{1}}\left(L_{-1}^{N}+\frac{1}{c}\sum_{k=2}^{N}\eta_{N,k}L_{-k}L_{-1}^{N-k}+\frac{1}{c^{2}}\sum_{\substack{k_{1},k_{2}=2\\
k_{1}\ge k_{2}
}
}^{N}\kappa_{N,k_{1},k_{2}}L_{-k_{1}}L_{-k_{2}}L_{-1}^{N-k_{1}-k_{2}}\right)\left|\mathcal{O}\right\rangle =0+\mathcal{O}\left(c^{-1}\right),
\end{equation}
\end{footnotesize}
with $m_{1},m_{2}\ge2$ and $m_{1}\ge m_{2}$, because $L_{m_{2}}L_{m_{1}}$
acting on the $\frac{1}{c^{2}}$ terms will contribute to leading
order $\mathcal{O}\left(c^{0}\right)$. 

We already know that 
\be
L_{m_{1}}\left(L_{-1}^{n}+\sum_{k=2}^{n}\lambda_{n,k}\frac{1}{c}L_{-k}L_{-1}^{n-k}\right)=0+\mathcal{O}\left(c^{-1}\right)
\ee
so in the following we only need to consider the remaining contribution
of the second term, which comes from $k=m_{2}$ and $k=m_{1}+m_{2}$,
\begin{small}
\begin{align}
L_{m_{2}}L_{m_{1}}\sum_{\substack{k=2\\
k\ne m_{1}
}
}^{N}\frac{1}{c}\lambda_{N,k}L_{-k}L_{-1}^{N-k}\left|\mathcal{O}\right\rangle = & \frac{m_{2}\left(m_{2}^{2}-1\right)}{12}\left[\right.\lambda_{N,m_{1}+m_{2}}\left(2m_{1}+m_{2}\right)\\
 & \left.+\lambda_{N,m_{2}}\frac{\left(N-m_{2}\right)!(h\left(m_{1}+1\right)+N-m_{2}-m_{1})}{\left(N-m_{2}-m_{1}\right)!}\right]L_{-1}^{N-m_{1}-m_{2}}\left|\mathcal{O}\right\rangle \nn
\end{align}
\end{small}
The third term in equation (\ref{eq:1OvercSquareCorrectionToPhi}) give the following leading order contribution 
\begin{align}
 & L_{m_{2}}L_{m_{1}}\sum_{\substack{k_{1},k_{2}=2\\
k_{1}\ge k_{2}
}
}^{N}\frac{1}{c^2}\kappa_{n,k_{1},k_{2}}L_{-k_{1}}L_{-k_{2}}L_{-1}^{n-k_{1}-k_{2}}\left|\mathcal{O}\right\rangle \\
= & \left(1+\delta_{m_{1},m_{2}}\right)\kappa_{N,m_{1},m_{2}}\frac{m_{1}\left(m_{1}^{2}-1\right)m_{2}\left(m_{2}^{2}-1\right)}{144}L_{-1}^{N-m_{1}-m_{2}}\left|\mathcal{O}\right\rangle +\mathcal{O}\left(c^{-1}\right).\nn
\end{align}
So equating the RHSs of the above two equations, and solving for $\kappa_{N,m_{1},m_{2}}$,
we find 
\begin{equation}\label{eq:KappaNm1m2}
\kappa_{N,m_{1},m_{2}}=-\frac{\lambda_{N,m_{1}+m_{2}}\left(2m_{1}+m_{2}\right)+\lambda_{N,m_{2}}\frac{\left(N-m_{2}\right)!\left(h\left(m_{1}+1\right)+N-m_{2}-m_{1}\right)}{\left(N-m_{2}-m_{1}\right)!}}{\left(1+\delta_{m_{1},m_{2}}\right)\frac{m_{1}\left(m_{1}^{2}-1\right)}{12}}.
\end{equation}
So $\mathcal{L}_{-N}$ is by given equation (\ref{eq:1OvercSquareInMathematicalL}) with $\eta_{N,k}$ and
$\kappa_{N,k_{1},k_{2}}$ given by equation (\ref{eq:EtaNm}) and equation (\ref{eq:KappaNm1m2}). 

Notice that the $\eta_{N,k}$ and $\kappa_{N,k_{1},k_{2}}$ we derived
above are just the leading order results, ie there are $\frac{1}{c}$
corrections to them. And there are other terms, like $L_{-1}^{N}$,
at order $\frac{1}{c}$ and $\frac{1}{c^{2}}$. In general, these
$\frac{1}{c}$ corrections should form quasi-primaries and their
global descendants, such that $\left\langle \phi\mathcal{O}\right\rangle $
will always be given by $\left\langle \phi\mathcal{O}\right\rangle =\left\langle \phi^{\text{global}}\mathcal{O}\right\rangle $,
which is just the bulk-boundary propagator in vacuum.

\subsection{Explicit Form of the Stress-Tensor Correlator Recursion and Calculation}
\label{app:underbraceRecursion}

We can document the origin of various terms in the recursion relation from section \ref{sec:RecursionRelationPhiOTT} very explicitly as
\begin{align*}
G_{n+1,m}= & \left(\underbrace{\underbrace{-\frac{\partial_{z}+\sum_{i=1}^{n}\partial_{z_{i}}}{z_{n+1}}}_{L_{-1}\phi}+\underbrace{\frac{\frac{y}{2}\partial_{y}}{z_{n+1}^{2}}}_{L_{0}\phi}\underbrace{\underbrace{-\frac{z\left(2h+z\partial_{z}\right)}{z_{n+1}^{3}}}_{-\left[L_{1},\mathcal{O}\left(z,\overline{z}\right)\right]}\underbrace{-\sum_{i=1}^{n}\frac{z_{i}\left(4+z_{i}\partial_{z_{i}}\right)}{z_{n+1}^{3}}}_{-\left[L_{1},T\left(Z_{i}\right)\right]}}_{L_{1}\phi}}_{T\left(z_{n+1}\right)\phi\left(y,0,0\right)}\right)G_{n,m}\\
 & +\left(\underbrace{\frac{h}{\left(z_{n+1}-z\right)^{2}}+\frac{\partial_{z}}{\left(z_{n+1}-z\right)}}_{T\left(z_{n+1}\right)\mathcal{O}\left(z,\overline{z}\right)}+\underbrace{\sum_{i=1}^{n}\left(\frac{2}{\left(z_{n+1}-z_{i}\right)^{2}}+\frac{\partial_{z_{i}}}{z_{n+1}-z_{i}}\right)}_{T\left(z_{n+1}\right)T\left(z_{i}\right)}\right)G_{n,m}\\
 & +\underbrace{\sum_{i=1}^{n}\frac{\left\langle T\left(z_{1}\right)\cdots T\left(z_{i-1}\right)T\left(z_{i+1}\right)\cdots T\left(z_{n}\right)\overline{T}\left(\overline{w}_{1}\right)\cdots\overline{T}\left(\overline{w}_{m}\right)\mathcal{O}\left(z,\overline{z}\right)\phi\left(y,0,0\right)\right\rangle }{2\left(z_{n+1}-z_{i}\right)^{4}}}_{T\left(z_{n+1}\right)T\left(z_{i}\right)}
\end{align*}
One can use the above recursion relation to easily derive $\left\langle \phi\mathcal{O}T\right\rangle $
, $\left\langle \phi\mathcal{O}TT\right\rangle $ and $\left\langle \phi\mathcal{O}T\overline{T}\right\rangle $
that we derived in section \ref{app:ComputationsBBOPEBlock} using bulk-boundary OPE block. For
comparison, we provide these computations here.

For one $T$ insertion, we have
\begin{align}
\left\langle \phi\left(y,0,0\right)\mathcal{O}\left(z,\overline{z}\right)T\left(z_{1}\right)\right\rangle = & \left(-\frac{\partial_{z}}{z_{1}}+\frac{\frac{y}{2}\partial_{y}}{z_{1}^{2}}-\frac{z\left(2h+z\partial_{z}\right)}{z_{1}^{3}}+\frac{h}{\left(z_{1}-z\right)^{2}}+\frac{\partial_{z}}{z_{1}-z}\right)\left\langle \phi\mathcal{O}\right\rangle \nn\\
= & \left(\frac{y}{z\bar{z}+y^{2}}\right)^{2h}\frac{hz^{2}\left(z_{1}\left(z\bar{z}+3y^{2}\right)-2y^{2}z\right)}{\left(z-z_{1}\right){}^{2}z_{1}^{3}\left(y^{2}+z\overline{z}\right)}.
\end{align}

For one $T$ and one $\overline{T}$ insertions, we have
\begin{align}
 & \left\langle \phi\left(y,0,0\right)\mathcal{O}\left(z,\overline{z}\right)T\left(z_{1}\right)\overline{T}\left(\overline{w}_{1}\right)\right\rangle\nn \\
= & \left(-\frac{\partial_{\overline{z}}}{\overline{w}_{1}}+\frac{\frac{y}{2}\partial_{y}}{\overline{w}_{1}^{2}}-\frac{\overline{z}\left(2h+\overline{z}\partial_{\overline{z}}\right)}{\overline{w}_{1}^{3}}+\frac{h}{\left(\overline{w}_{1}-\overline{z}\right)^{2}}+\frac{\partial_{\overline{z}}}{\overline{w}_{1}-\overline{z}}\right)\left\langle \phi\left(y,0,0\right)\mathcal{O}\left(z,\overline{z}\right)T\left(z_{1}\right)\right\rangle\nn \\
= & \left(\frac{y}{z\bar{z}+y^{2}}\right)^{2h}\left(\frac{h^{2}z^{2}\bar{z}^{2}\left(y^{2}\left(3\bar{w}_{1}-2\bar{z}\right)+\bar{w}_{1}z\bar{z}\right)\left(y^{2}\left(3z_{1}-2z\right)+z_{1}z\bar{z}\right)}{z_{1}^{3}\overline{w}_{1}^{3}\left(z_{1}-z\right){}^{2}\left(\overline{w}_{1}-\bar{z}\right)^{2}\left(z\bar{z}+y^{2}\right)^{2}}\right.\\
 & \qquad\qquad\qquad\  \left.+\frac{2hy^{2}z^{3}\bar{z}^{3}}{z_{1}^{3}\bar{w}_{1}^{3}\left(z-z_{1}\right)\left(\bar{w}_{1}-\bar{z}\right)\left(z\bar{z}+y^{2}\right)^{2}}\right).\nn
\end{align}

For two $T$ insertions, we have
\begin{small}
\begin{align}
 & \left\langle \phi\left(y,0,0\right)\mathcal{O}\left(z,\overline{z}\right)T\left(z_{1}\right)T\left(z_{2}\right)\right\rangle \nn\\
= & \left(-\frac{\partial_{z}+\partial_{z_{1}}}{z_{2}}+\frac{\frac{y}{2}\partial_{y}}{z_{2}^{2}}-\frac{z\left(2h+z\partial_{z}\right)}{z_{2}^{3}}-\frac{z_{1}\left(4+z_{1}\partial_{z_{1}}\right)}{z_{2}^{3}}\right)\left\langle \phi\left(y,0,0\right)\mathcal{O}\left(z,\overline{z}\right)T\left(z_{1}\right)\right\rangle \nn\\
 & +\left(\frac{h}{\left(z_{2}-z\right)^{2}}+\frac{\partial_{z}}{z_{2}-z}+\frac{2}{\left(z_{2}-z_{1}\right)^{2}}+\frac{\partial_{z_{1}}}{z_{2}-z_{1}}\right)\left\langle \phi\left(y,0,0\right)\mathcal{O}\left(z,\overline{z}\right)T\left(z_{1}\right)\right\rangle \nn\\
 & +\frac{c}{2\left(z_{2}-z_{1}\right)^{2}}\left\langle \phi\left(y,0,0\right)\mathcal{O}\left(z,\overline{z}\right)\right\rangle \\
= & \left(\frac{y}{z\bar{z}+y^{2}}\right)^{2h}\left[\frac{c}{2\left(z_{1}-z_{2}\right){}^{4}}+\frac{h^{2}z^{4}\left(z_{1}z\bar{z}+y^{2}\left(3z_{1}-2z\right)\right)\left(z_{2}z\bar{z}+y^{2}\left(3z_{2}-2z\right)\right)}{z_{1}^{3}z_{2}^{3}\left(z-z_{1}\right){}^{2}\left(z-z_{2}\right)^{2}\left(z\bar{z}+y^{2}\right)^{2}}\right.\nn\\
 & \left.+\frac{2hz^{2}\left(y^{2}z\bar{z}z_{1}z_{2}\left(z\left(z_{1}+z_{2}\right)-4z_{1}z_{2}\right)-z^{2}\bar{z}^{2}z_{1}^{2}z_{2}^{2}+y^{4}\left(zz_{1}z_{2}\left(z_{1}+z_{2}\right)-3z_{1}^{2}z_{2}^{2}-z^{2}\left(z_{1}-z_{2}\right)^{2}\right)\right)}{\left(z-z_{1}\right)z_{1}^{3}z_{2}^{3}\left(z_{2}-z\right)\left(z_{2}-z_{1}\right){}^{2}\left(z\bar{z}+y^{2}\right)^{2}}\right].\nn
\end{align}
\end{small}
One can see that the above results are exactly what we found in section \ref{app:ComputationsBBOPEBlock}.

\bibliographystyle{utphys}
\bibliography{VirasoroBib}

\end{document}